\documentclass[aps,prd,twocolumn,nofootinbib,superscriptaddress]{revtex4}
\usepackage{latexsym}
\usepackage{hyperref}
\usepackage{amssymb}
\usepackage{epsfig,amsmath,graphics}
\usepackage{pstricks}
\usepackage{color}
\usepackage{dcolumn}
\usepackage{ulem}
\usepackage{xcolor}
\usepackage{placeins}

\definecolor{linkcolor}{rgb}{0.0, 0.28, 0.67}
\hypersetup{colorlinks=true,
linkcolor=black,
citecolor=linkcolor,
urlcolor=linkcolor,
linktocpage=true
}

\newcommand{\tev}{\text{TeV}}

\newcommand{\mpc}{\text{Mpc}}

\newcommand{\D}{\cal{D}}

\newcommand{\QDMAP}{Q_\text{DMAP}}

\newcommand{\nir}{N_\text{IR}}
\newcommand{\nuv}{N_\text{UV}}

\usepackage{comment}
\newcommand{\be}{\begin{equation}}
\newcommand{\ee}{\end{equation}}

\def\bea{\begin{eqnarray}}
\def\eea{\end{eqnarray}}
\def\ltap{\ \raise.3ex\hbox{$<$\kern-.75em\lower1ex\hbox{$\sim$}}\ }
\def\gtap{\ \raise.3ex\hbox{$>$\kern-.75em\lower1ex\hbox{$\sim$}}\ }
\def\lsim{\ \raise.3ex\hbox{$<$\kern-.75em\lower1ex\hbox{$\sim$}}\ }
\def\gsim{\ \raise.3ex\hbox{$>$\kern-.75em\lower1ex\hbox{$\sim$}}\ }

\usepackage{slashed}

\usepackage{color}

\newcommand{\zt}{z_t}

\newcommand{\ignore}[1]{}
\newcommand{\beq}{\begin{equation}}
\newcommand{\eeq}{\end{equation}}
\newcommand{\bear}{\begin{eqnarray}}
\newcommand{\eear}{\end{eqnarray}}

\def\tev{\,{\rm TeV}}

\def\ev{\,{\rm eV}}
\def\lcdm{$\Lambda{\rm CDM}$}

\newcommand{\Neff}{N_{\rm eff}}

\newcommand{\LCDM}{\Lambda{\rm CDM}}

\newcommand{\dataD}{$\mathcal{D}\,$}
\newcommand{\dataDH}{$\mathcal{DH}\,$}
\newcommand{\dataDS}{$\mathcal{DS}\,$}
\newcommand{\dataDHS}{$\mathcal{DHS}\,$}

\begin{document}

\title{A Step in Understanding the $S_8$ Tension}

\author{Melissa Joseph}\thanks{These authors contributed equally to this work.}
\affiliation{Physics Department, Boston University, Boston, MA 02215, USA}
\author{Daniel Aloni}\thanks{These authors contributed equally to this work.}
\affiliation{Physics Department, Boston University, Boston, MA 02215, USA}
\author{Martin Schmaltz}
\affiliation{Physics Department, Boston University, Boston, MA 02215, USA}
\author{Eashwar N. Sivarajan}
\affiliation{Physics Department, Boston University, Boston, MA 02215, USA}
\author{Neal Weiner}
\affiliation{Center for Cosmology and Particle Physics, Department of Physics, New York University, New York, NY 10003, USA}
\begin{abstract}
Models of dark sectors with a mass threshold can have important cosmological signatures. If, in the era prior to recombination, a relativistic species becomes nonrelativistic and is then depopulated in equilibrium, there can be measurable impacts on the cosmic microwave background as the entropy is transferred to lighter relativistic particles. In particular, if this ``step'' occurs near $z\sim 20,000$, the model can naturally accommodate larger values of $H_0$. If this stepped radiation is additionally coupled to dark matter, there can be a meaningful impact on the matter power spectrum as dark matter can be coupled via a species that becomes nonrelativistic and depleted. This can naturally lead to suppressed power at scales inside the sound horizon before the step, while leaving conventional cold dark matter signatures for power outside the sound horizon. We study these effects and show such models can naturally provide lower values of $S_8$ than scenarios without a step. This suggests these models may provide an interesting framework to address the $S_8$ tension, both in concert with the $H_0$ tension and without.
\end{abstract}

\maketitle

\section{Introduction}\label{sec: Introduction}
The past two decades have seen cosmology become a precision science. A wealth of new data over wide ranges of redshifts and distance scales has appeared, allowing a precision determination of the parameters of the concordance model ($\Lambda$CDM). In addition, as error bars continue to shrink, we now look forward to an era where even small deviations from $\Lambda$CDM might show themselves as a statistically significant deviation in the data.

There are many reasons to expect that such deviations should appear at some point. Most models of cold dark matter (CDM) have interactions at some level, with itself and with other species, possibly within the Standard Model. Additional dark radiation (DR) is naturally populated by thermal contact in the early Universe, although it may be proportionately small today due to the large number of degrees of the Standard Model, or some other source of entropy. 

While many properties of high-energy physics are hidden in the smallest scales, many different observables are sensitive to physical processes from when the photon bath had a temperature of a keV and below. Although this is much smaller than many particle physics scales, it is larger than other important scales known in nature. In particular, both the neutrino mass ($m_\nu \sim 0.1 \ev$) and the cosmological constant scale [$\Lambda^4 \sim (10^{-2.5} \ev)^4$] are physical scales below this. In particle physics models, there are frequently mass scales induced at scales in the several orders of magnitude around $\sim \tev^2/M_{pl} \sim 10^{-4} \ev$. Thus, not only can cosmology probe the presence of additional sectors of physics, presently disconnected from ours, it probes a potentially interesting energy range, as well.

Recently, the consequences of a step in a fluid of self-interacting dark radiation (SIDR) was investigated in the context of the $H_0$ tension \cite{Aloni:2021eaq}. The $H_0$ tension is the $4.8\sigma$ disagreement between the measurement of the local expansion rate based on late-Universe observables, in particular Cepheid-calibrated Type IA supernovae from SH0ES \cite{Riess:2021jrx} $H_0=73.04\pm1.04\,$km/s/Mpc when compared against inferences based early-Universe physics, such as from the cosmic microwave background (CMB) alone or in combination with Baryon acoustic oscillations (BAO) $H_0=67.27\pm0.60\,$km/s/Mpc~\cite{Planck:2018vyg}. A broad variety of proposals have been presented in attempts to reconcile this, see., e.g., \cite{Brust:2017nmv,Escudero:2019gvw,EscuderoAbenza:2020egd,Escudero:2021rfi,Karwal:2016vyq,Poulin:2018cxd,Lin:2019qug,Smith:2019ihp,Cyr-Racine:2021alc,Berghaus:2022cwf,Niedermann:2020dwg} and additional models summarized in \cite{2203.06142}. One common group of models include additional radiation \cite{Buen-Abad:2015ova,Lesgourgues:2015wza,Buen-Abad:2017gxg,Bernal:2016gxb, Blinov:2020hmc, RoyChoudhury:2020dmd,Brinckmann:2020bcn,Kreisch:2019yzn,Chacko:2016kgg,Choi:2020pyy,Bansal:2021dfh}, which may be strongly interacting. In \cite{Aloni:2021eaq}, it was shown that the presence of a mass threshold in a fluid of SIDR allowed for a higher value of $H_0$ and a better overall fit to a broad set of data. In particular, the data preferred a mass threshold occurring near $z_t \sim 20\,000$.

The  model considered there was a simple two-component model, with a massive scalar and a massless fermion. This simple model is neatly packaged in the simplest possible supersymmetric model, and was thus termed Wess-Zumino dark radiation (WZDR). As the scalar becomes nonrelativistic, its entropy is transferred to the fermion, heating it, and increasing the effective radiation compared to the step-less case. It is clear that extensions of this model are interesting to study. Probably the most immediately obvious to consider is the question; what if the dark matter {\it additionally} interacts with a portion of this dark fluid, such as, e.g., a Yukawa coupling to the scalar?

Naively, this extension would lead to a suppression of matter power for modes that were inside the horizon while the dark matter is coupled to the dark radiation fluid. In contrast, modes that come inside the horizon later would not be suppressed, leading to CDM phenomenology on large scales, and deviations at smaller scales. The comoving horizon for modes entering at redshift $z$ before matter radiation equality is approximately $r_{s} \sim 100\, \mpc\, {z_{eq}}/{z}$. Thus, for the parameters which are preferred by the $H_0$ tension, we expect deviations from CDM on scales of $\sim 10\, h^{-1}\mpc$ and below. Such a scale is interesting in light of a separate tension, namely the $S_8$ tension.

Loosely speaking, the $S_8$ tension is the difference between more direct measurements of the level of fluctuations in the matter power spectrum on $\sim$8 h$^{-1}$Mpc scales and their prediction using $\Lambda$CDM and the CMB to normalize. More quantitatively, the directly measured values from KiDS-1000 \cite{Heymans:2020gsg} and DES Y3 \cite{DES:2021wwk} combined give $S_8 \equiv \sqrt{\Omega_m/0.3}\, \sigma_8 = 0.769 \pm 0.016$ which is $2.9\sigma$ lower than, for example, the value obtained from the CMB by Planck \cite{Planck:2018vyg} $S_8=0.834\pm0.016$. Although this tension is not as statistically significant as the $H_0$ tension, it has appeared across many different independent measurements of $S_8$  \cite{2203.06142}. Moreover, it raises a basic questions as to what sorts of models might impact the power spectrum at measurable scales, while leaving CDM-like cosmology on larger scales. The fact that a simple model extension immediately points to consequences relevant at the appropriate scale is a striking coincidence. 

In this paper we will explore the consequences of dark matter interacting with a stepped fluid, where the late-time behavior after the step of dark matter is that of CDM. The layout of the paper is as follows: in Sec.~\ref{sec:model} we lay out the specific scenarios we wish to consider, namely the WZDR model of \cite{Aloni:2021eaq} where the dark matter additionally interacts with the scalar. We study the cosmological signals of such a scenario in Sec.~\ref{sec:cosmosignals}, showing that such scenarios naturally suppress matter power on $\sim 15 \, \mpc \times [20\,000/z_{t}]$ scales and smaller. We perform a global fit in Sec.~\ref{sec:fits} to an extended dataset including a wide variety of cosmological datasets, finding a good improvement in fits compared to $\Lambda$CDM and WZDR with CDM. Finally, in Sec.~\ref{sec:conclusions}, we conclude. 
Supplementary information is given in the appendices. Appendix~\ref{app: acronyms} contains a concise description of the models we study. Details regarding the DM-DR interaction and its implementation in the Boltzmann solver are given in Appendix~\ref{app: momentum transfer rate}. We discuss some issues regarding nonlinearities in Appendix~\ref{app: halofit}. Finally a full set of posterior densities and best-fit cosmological parameters as determined by our Markov chain Monte Carlo (MCMC) analysis, are provided in Appendix~\ref{app: triangle plots and tables}.

\section{Model of a Stepped Dark Sector}\label{sec:model}
The Wess-Zumino dark radiation (WZDR) model~\cite{Aloni:2021eaq} is a  simple and natural example of a dark sector with a mass threshold. It contains just two particles -- a fermion $\psi$ and a scalar $\phi$, which interact through a Yukawa coupling $\phi \psi \psi$. If we allow for scalar quartic interactions $\phi^4$, this model is efficiently packaged into the simplest known supersymmetric model, namely, the Wess-Zumino model. 
Supersymmetry breaking induces a small mass $m_\phi^2 \phi^2$ for the scalar, which can naturally lie near the \ev\, scale. Importantly, the dynamics we discuss are independent of supersymmetry, although this provides natural model-building directions. 

At early times, before the CMB era, some process produces \footnote{We assume that this process  occurs after big bang nucleosynthesis (BBN) in order to preserve its successful prediction of the helium abundance.} the WZDR $\psi$ and $\phi$ particles with an energy density equivalent to $\nuv$ additional neutrino species. 
As $\phi$ and $\psi$ always maintain chemical and kinetic equilibrium, 
once the temperature of this sector decreases below $m_\phi$, the scalars decay and annihilate, 
depositing their entropy into the lighter $\psi$ species.
Due to this process which takes approximately a decade in redshift, the relative energy density of the fluid, 
as quantified by the effective number of additional neutrinos species $N (z)$, increases to a value $\nir  = (15/7)^{1/3} \nuv$. 
Assuming a temperature of the dark radiation today of $T_{d0}$, the transition approximately starts at redshift $1+z_t = m_\phi / T_{d0}$, 
and the evolution of $N(z)$ as a function of redshift is calculated by solving the entropy conservation equation. 

This simple model was shown to significantly alleviate the Hubble tension, with the introduction of a single low-mass threshold, a ``step''~\cite{Aloni:2021eaq}. It is easy to imagine extensions of this model, perhaps the simplest of which is if the relativistic fluid additionally has interactions with dark matter, $\chi$.  If $\chi$ is a fermion, the simplest interaction possible is one in which an additional Yukawa coupling is added $\phi \chi \chi$. Interactions with the fluid would arise through Compton-like $\phi \chi \rightarrow \phi \chi$ processes, as well as $\chi \psi \rightarrow \chi \psi$ mediated by $t$-channel $\phi$ exchange.  

Of these, the Compton-like process is typically smaller, suffering an additional $T_d^2/M_\chi^2$ suppression.  
At high temperatures, $T_d \gtrsim m_\phi$, the momentum transfer rate between the DM and the WZDR sector from the $t$-channel process scales as~\cite{Buen-Abad:2015ova} 
\begin{eqnarray}
	\Gamma \propto \frac{T_d^2 }{M_\chi} \qquad \text{for }T_d \gtrsim m_\phi~, 
\end{eqnarray}
where $M_\chi$ is the mass of the DM. Therefore during radiation domination, the momentum transfer rate scales as Hubble, and the ratio $\Gamma/H$  neither increases nor decreases over time. 

However, at late times, once the temperature drops below $m_\phi$, the $\psi-DM$ interaction is effectively given by the four-Fermi contact operator, $\psi^2 \chi^2/m_\phi^2$ which gives a suppressed momentum transfer rate that scales as
\begin{eqnarray}
	\Gamma \propto \frac{T_d^2 }{M_\chi}\left(\frac{T_d}{m_\phi}\right)^4 \qquad \text{for }T_d \lesssim m_\phi~,
\end{eqnarray}
and the interaction shuts off quite rapidly after the transition time $z_t$.

We found that to a very good approximation the momentum transfer rate is given by the phenomenologically motivated fitting formula
\begin{align}\label{eq: approximate momentum transfer rate}
	\Gamma(x) & = \Gamma_0\frac{(1+z_t)^2}{x^2} \left(\frac{1}{1 - 0.05 \sqrt{x} + 0.131 x }\right)^4 ~,
\end{align}
where $x \equiv m_\phi / T_d$ and $\Gamma_0$ is the momentum transfer rate extrapolated to today in a theory in which there is no step (i.e. where $m_\phi=0$). The coefficients have been chosen to best approximate the  exact numerical result. A full derivation of the momentum transfer rate is given in Appendix~\ref{app: momentum transfer rate}. From now on we will refer to a WZDR model with dark matter -- dark radiation interaction that shuts off as Eq.~\eqref{eq: approximate momentum transfer rate} as WZDR+.

As a consequence, we have a scenario in which at early times dark matter exchanges momentum with a relativistic fluid made up of $\psi$ and $\phi$ particles. This momentum transfer acts as a mild friction on the growth of matter perturbations. At late times, after $z_t$, we have a CDM-like $\chi$ decoupled from a still tightly self-coupled fluid of $\psi$. This mass threshold then naturally produces a CDM-like cosmology at late times, and a very different one beforehand.

\section{Effects of a Stepped Dark Sector}\label{sec:cosmosignals}
Stepped fluids which interact with the dark matter have interesting phenomenology, 
and variety of effects and imprints during the evolution of the Universe. Here we discuss some of them, 
emphasizing that there are two distinct effects which are both sensitive to the WZDR mass scale. 

The first effect which was discussed in detail in Ref.~\cite{Aloni:2021eaq} is due to the increase in the effective number of degrees of freedom.
Through the transition, as the energy density increases, the expansion rate of the Universe increases accordingly. 
As a result, there is an $\ell$-dependent phase shift of the CMB power spectrum compared to a model of interacting radiation without a step.
To leading order there is no phase shift for small $k$-modes that enter the horizon after the transition, 
while there is a linear phase-shift for large $k$-modes that enter the horizon before the transition. 
Therefore, as the change in the behavior of the phase shift depends on the time of transition,
the CMB is sensitive to $m_\phi$ or equivalently to $z_t$. 
Ref.~\cite{Aloni:2021eaq} found that this shift is strongly preferred by the full dataset including measurements of $H_0$ and allows for better fits with higher expansion rates.

The new ingredient of the model which we introduce here is the interaction of the stepped fluid with the dark matter. 
During the time of radiation domination and well before the transition the momentum transfer rate scales as Hubble, $\Gamma \propto H$, and therefore remains equally important throughout this period. 
Since in our model the WZDR fluid interacts with $100\%$ of the DM, similar to the study of~\cite{Buen-Abad:2015ova,Buen-Abad:2017gxg},
the WZDR-DM coupling is weak and the momentum transfer rate is always smaller then Hubble. 
As a result the DM does not  oscillate but only experiences a friction as it falls into the gravitational potentials.

As explained in Ref.~\cite{Buen-Abad:2017gxg}, the effect on the matter power spectrum  (MPS)
compared to a model with no interaction between the fluid and the DM, is a linear suppression in $\log k$ space 
with a slope proportional to the momentum transfer rate,
\begin{align*}
	\frac{P_{\text{ interacting}}}{P_{\text{not-interacting}}} & \simeq 
	\begin{cases}
		1 & k \ll k_{s.o.} \\
		1 - \sqrt{2}\,  \Gamma\!/\!H\times
		\log k / k_{s.o.}& k \gg k_{s.o.}
	\end{cases}~.
\end{align*}

\begin{figure}[t]
	\centering
	\includegraphics[width=0.45\textwidth]{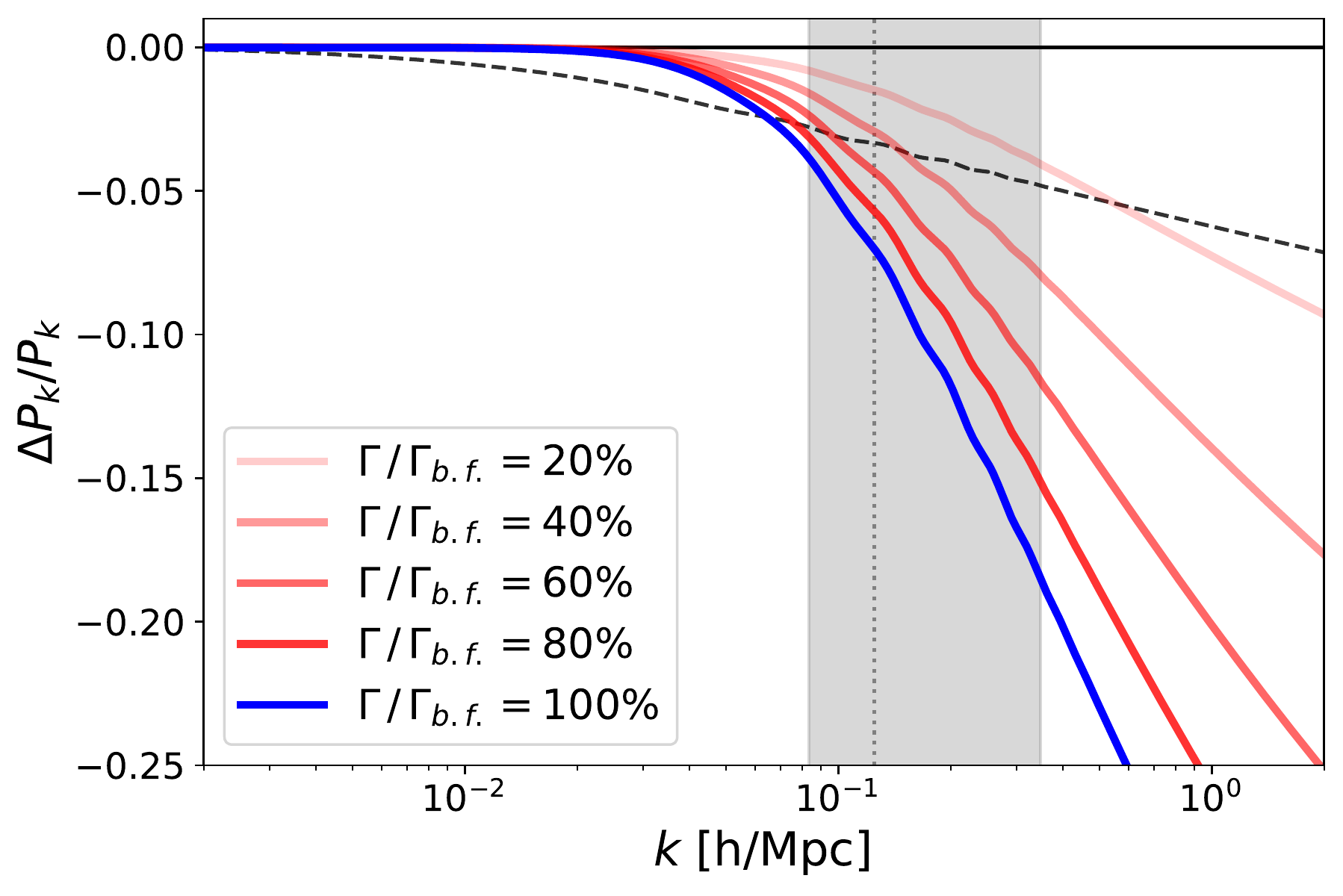}  \\
	\includegraphics[width=0.45\textwidth]{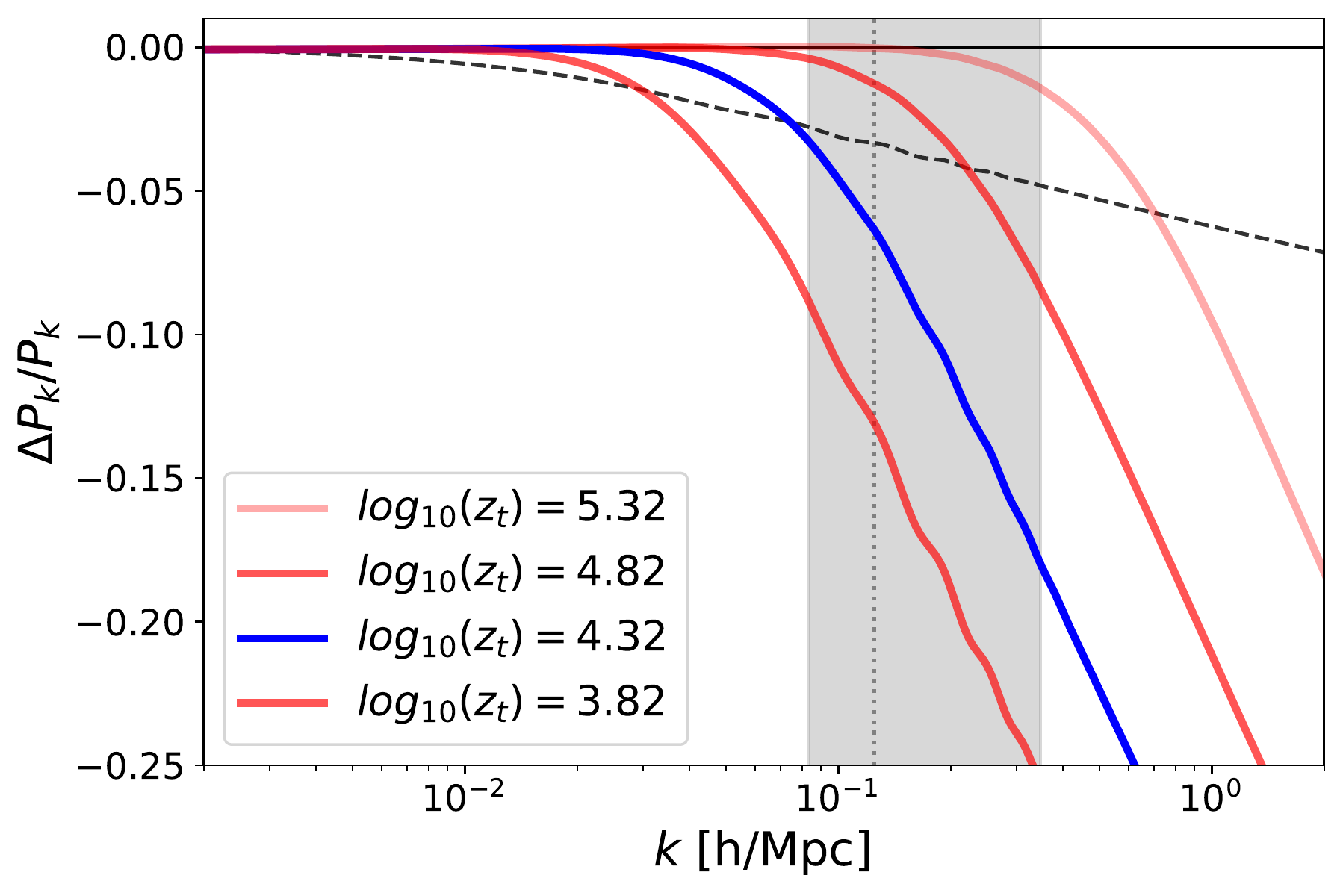}
	\caption{Dependence of the MPS on model parameters. Both panels show ratios of the MPS in WZDR+ compared to a reference model with no DR-DM interaction. The blue line indicates the best-fit point from our $\mathcal{DHS}$ fit, with $10^7 \Gamma_{b.f} = 5\, {\rm Mpc}^{-1}$. In the top panel the different lines correspond to different interaction strengths $\Gamma$ compared to the reference model with no interaction (i.e. WZDR). The bottom panel shows how the  MPS varies with the redshift of the transition $z_t$ compared to a reference model with $z_t\to \infty$. The black dashed line in both plots is the best-fit of SIDR+ to the \dataDHS fit compared to the same parameter space point with the DM-DR interactions shut off. The gray band shows the scales to which $S_8$ is most sensitive to through the $\sigma_8$ window function and the dotted gray line indicates $k=1/8\, h$Mpc$^{-1}$.}
	\label{fig:Pk effects}
\end{figure}

Here  $k_{s.o.}$ is the wave number of the mode which enters the horizon when the interaction shuts off.
This effect is shown in the top panel of Fig.~\ref{fig:Pk effects}, where the linear suppression is evident
and the slope is proportional to the momentum transfer rate. The fact that this suppression is smooth in $\log k$
and does not introduce a sharp feature or drop-off in the MPS allows this kind of model to
lower $S_8$ and fit the MPS extracted from Lyman-$\alpha$ data which prefer a steeper
slope at the scale $k\sim 1$ Mpc$^{-1}$~\cite{Pan:2018zha}. 

In Ref.~\cite{Buen-Abad:2017gxg} the interaction becomes smaller compared to Hubble after matter-radiation equality, as $H\propto a^{-3/2}$
while the interaction still drops as $\Gamma\propto a^{-2}$, thus the shut-off time is at the time of equality $\eta_{eq.}$
In contrast, for the WZDR+ model the momentum transfer rate shuts off once the temperature drops below $m_\phi$ (see Eq.~\eqref{eq: approximate momentum transfer rate}). This is shown in the bottom panel of Fig.~\ref{fig:Pk effects}. 
As a result the matter power spectrum, through the $S_8$ measurements, is sensitive to $m_\phi$ or equivalently to $z_t$.

\section{Analysis}\label{sec:fits}
Within the WZDR+ framework, there are some immediate questions; namely, can this model give a good description of existing data? In particular, can it address the known tensions in the data? And finally, is there consistency between the different datasets in the value of $z_t$ extracted? In this section we consider precisely these questions. We will study how well WZDR+ can improve the overall fit to the CMB and alleviate the Hubble and $S_8$ tensions. Here we highlight our main results, more details from our analysis can be found in Appendices~\ref{app: halofit} and~\ref{app: triangle plots and tables}. 

We modified CLASS v3.1~\cite{Blas:2011rf} to include the stepped fluid~\cite{Aloni:2021eaq} and further modified the code to include interactions with DM as described in Appendix~\ref{app: momentum transfer rate}. 
We use the MontePython v3.5~\cite{Brinckmann:2018cvx, Audren:2012wb} MCMC sampler to study the constraints of various datasets on the model.
Similar to Ref.~\cite{Aloni:2021eaq}, for the WZDR+ model, we adopt flat priors on the six $\LCDM$ cosmological parameters $\{\omega_b,\, \omega_{\rm dm},\, \theta_s, n_s, A_s,\, \tau_{\rm reio} \}$. For the 3 new parameters of WZDR+ we include a flat prior on the amount of dark radiation after the step\footnote{The lower bound was included to avoid numerical issues of our code near $\nir=0$.  We explicitly checked that our results are not very sensitive to small changes of this bound.} $\nir > 0.01$, a logarithmic prior on the redshift of the step location\footnote{These bounds are designed to avoid scanning over models in which the transition occurs too early or too late to have much effect on the CMB, see~\cite{Aloni:2021eaq}.}  $\log_{10} (z_t) \in  [4.0, 4.6]$, and a linear prior on the strength of the interaction between dark radiation and dark matter $\Gamma_0>0$.
Finally, we assume that the extra radiation in WZDR+ (and SIDR+, see below) is populated after BBN so that the predicted abundance of primordial helium $Y_p$ is sensitive only to the Standard Model radiation at BBN ($ \Neff = 3.044.$).

We consider combinations of three datasets:
\begin{itemize}
	\item Our baseline dataset $\mathcal{D}$ includes the Planck 2018~\cite{Planck:2018vyg}, TT,TE, and EE  data for low-$\ell$ (`lowl\_TT', `lowl\_EE') and high-$\ell$ (`highl\_TTTEEE') with the full set of nuisance parameters. 
	It also includes the late-Universe constraints: the BAO-only likelihood (`bao\_boss\_dr12') from BOSS DR12 ($z = 0.38, 0.51, 0.61$)\cite{arxiv:1607.03155} and the small-z BAO likelihood (`bao\_smallz\_2014') including data from the 6dF ($z = 0.106$)\cite{arXiv:1106.3366} and MGS ($z = 0.15$)~\cite{arXiv:1409.3242} catalogs, as well as the PANTHEON~\cite{1710.00845} supernova likelihood (`Pantheon').
	\item The dataset $\mathcal{H}$ is chosen to test the Hubble tension, it consists of the latest measurement of the intrinsic magnitude of supernovae $M_b=-19.253\pm0.027$ by the SHOES Collaboration~\cite{Riess:2021jrx} , which we implement as a Gaussian likelihood for this parameter.  
	\item The dataset $\mathcal{S}$ is chosen to test the $S_8$ tension. It includes the $3\times 2 pt$ weak lensing and galaxy clustering analyses by KiDS-1000$\times$\{2dFLenS+BOSS\}~\cite{Heymans:2020gsg} and DES-Y3~\cite{DES:2021wwk} which obtain $S_8=0.766^{+0.020}_{-0.014}$ and $S_8=0.775^{+0.026}_{-0.024}$, which we implement as simple asymmetric Gaussian likelihoods for $S_8$. For quantifying the ``Gaussian tension'' we also combine the two $S_8$ measurements and their positive $1\sigma$ ranges to $S_8^{\rm direct} = 0.769 \pm 0.016$. In addition, the dataset $S$ includes the Planck lensing~\cite{Planck:2018vyg} likelihood (`Planck\_lensing'). 
\end{itemize}

A potential concern regarding our fit to data for the BAO scale and $S_8$ is that the experimental values for these parameters have been obtained from observations which assumed $\Lambda$CDM templates for the fit to data.
The BAO scale is well-known to be robust against smooth changes to the MPS from new physics such as the ones predicted in our model and also against nonlinear corrections to the MPS~\cite{Bernal:2020vbb}. A separate concern, highlighted in \cite{Bernal:2020vbb}, is that a $\Lambda$CDM-based extraction of the BAO scale can be biased for models which induce phase shifts of the BAO peaks from perturbations. In the DNI model studied in \cite{Bernal:2020vbb}, strong neutrino-DM interactions suppress free-streaming of the SM neutrinos which reduces the well-known BAO phase shift originating from the free-streaming neutrinos. This reduced phase shift is not accounted for in the $\Lambda$CDM-based analysis and leads to a small bias in the extracted BAO scale. A similar phase shift might naively be expected in WZDR+ which has interactions between the DR and DM. However, the interactions between the DM and DR rapidly shuts off well before recombination ($z_t\sim 2 \times 10^4$), and therefore have minimal impact on the physics of BAO. In addition, the SM neutrinos in WZDR+ free-stream as in $\Lambda$CDM while the small amount of DR remains self-interacting throughout. Thus the phase shifts associated with the DM-DR interactions as in DNI do not arise in WZDR+.

The main concern with $S_8$ is that it is extracted from scales at which nonlinear corrections to the matter power spectrum are not negligible, and one could worry that the nonlinear corrections applied by the experiments produce biased results for parameter estimations in the presence of physics beyond $\Lambda$CDM. 

To alleviate this concern we show in Appendix~\ref{app: halofit} that for the range of $k$-modes which are significant for the extraction of $S_8$ the WZDR+ linear MPS is well-approximated by the MPS of a $\Lambda$CDM ``Avatar'' with suitably adjusted $n_s$ and $A_s$. Thus for the relevant range of $k$ the nonlinear growth of structure in our model is expected to be identical to that in $\Lambda$CDM\footnote{As pointed out by~\cite{Amon:2022azi} the KiDS-1000 result might be explained within $\Lambda$CDM by unexpectedly large baryonic feedback effects. We do not address this issue here but show that no significant additional modifications from nonlinear effects are expected in the WZDR+ model compared to $\Lambda$CDM.} (the DM-DR interactions turn off before matter-radiation equality). To explicitly demonstrate this, we plot the nonlinear corrections for WZDR+ and the Avatar obtained from CLASS with {\it HMcode}~\cite{Mead:2015yca} and {\it halofit}~\cite{Smith:2002dz,Takahashi:2012em} in Appendix~\ref{app: halofit} and show that they are well-behaved and agree as well as they do in standard $\Lambda$CDM. We conclude that we do not expect the nonlinear corrections to the MPS in WZDR+ to bias the extraction of $S_8$ significantly.

We do not include the ACT DR4~\cite{ACT:2020gnv} and SPT 3G~\cite{SPT-3G:2022hvq} datasets. Although these high-resolution CMB experiments promise great sensitivity in the future, today ACT and SPT data are still less constraining than Planck 2018, and there are known tensions between the ACT and Planck data which would warrant a more careful analysis. Given the early stage of the ACT and SPT data and the promise of very significant improvements, we defer such an  analysis to a future study.
We will perform fits to the combinations of datasets $\mathcal{D}$, $\mathcal{DH}$, $\mathcal{DS}$, and $\mathcal{DHS}$.

To put our fits in perspective we compare WZDR+ to two other models: $\Lambda$CDM and SIDR+, a model in which self-interacting dark radiation (SIDR) weakly interacts with the DM~\cite{Buen-Abad:2015ova,Buen-Abad:2017gxg}. SIDR+ is a natural model to compare to because it also has {(i)} extra radiation which is important for the Hubble tension and {(ii)} friction between dark radiation and dark matter which is important to address the $S_8$ tension; relative to WZDR+ it is just missing the mass threshold, the ``step''.
\begin{figure}[t]
	\includegraphics[width=0.45\textwidth]{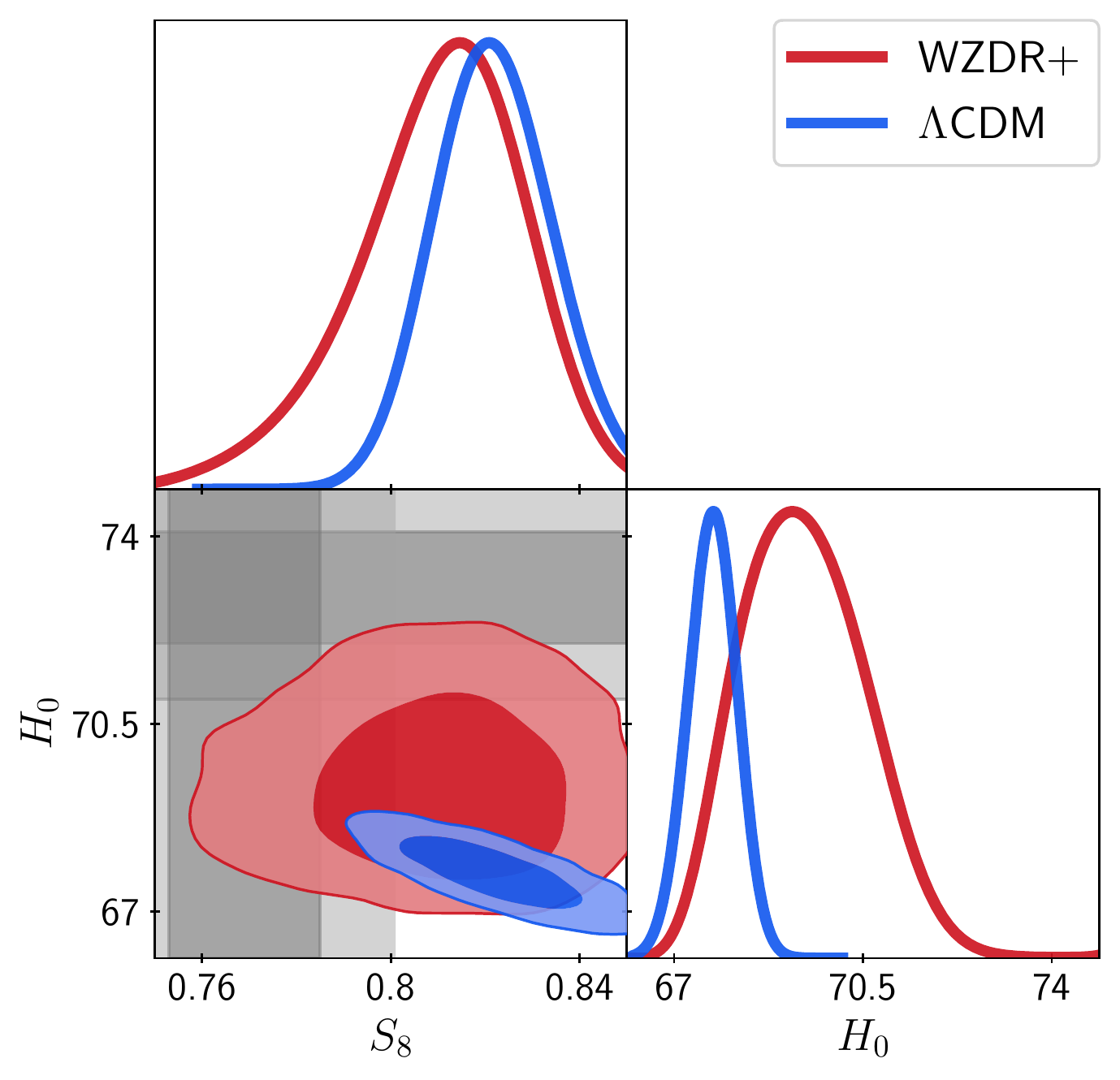}
	\caption{Posterior distributions for $\Lambda$CDM vs. WZDR+ fitted to the dataset $\mathcal{D}$. The contours of WZDR+ are much broader and in a better consistency with direct measurements of $S_8$ and $H_0$. The gray bands show the one and two sigma regions of $H_0$ from the SH0ES Collaboration, and of our combination $S_8^{\rm direct}$.}
	\label{fig:wzdrPlus_vs_LCDM_D}
\end{figure}

\begin{figure}[t]
	\includegraphics[width=0.45\textwidth]{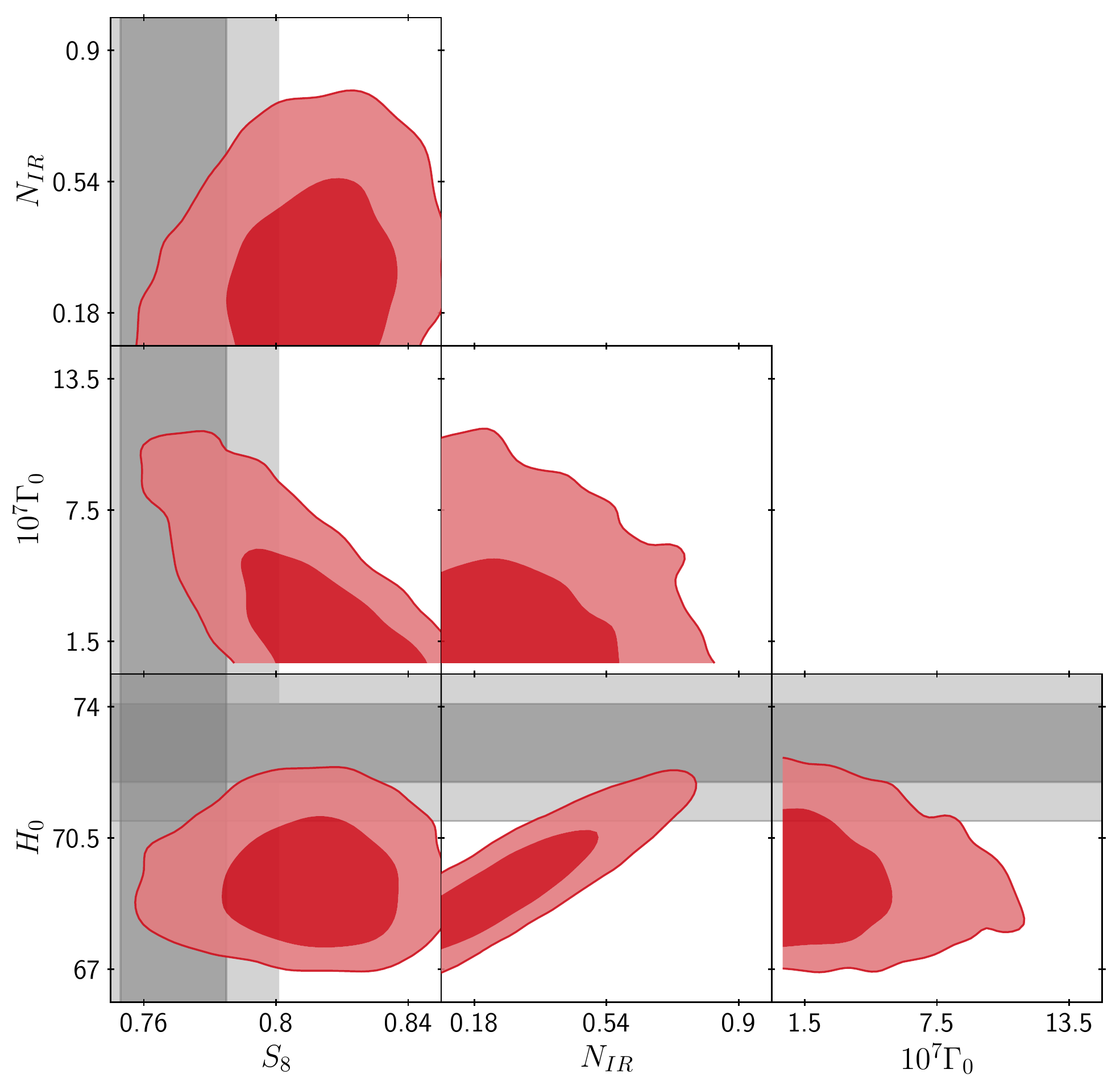}
	\caption{Posterior distribution of WZDR+ fitted to the $\mathcal{D}$ dataset. The distribution shows a clear correlation between the model parameters $N_{IR}$ and $\Gamma_0$ and the inferred quantities $H_0$ and $S_8$. The gray bands show the one and two sigma regions of $H_0$ from the SH0ES Collaboration, and of our combination $S_8^{\rm direct}$.}
	\label{fig:wzdrPlus_D_triangle}
\end{figure}

A first question to address is whether WZDR+ can provide an overall better fit to the data, and, in particular, whether it ameliorates both the $H_0$ and $S_8$ tensions simultaneously. Fig.~\ref{fig:wzdrPlus_vs_LCDM_D} shows the posterior in the $H_0-S_8$ plane of our fit of WZDR+ to the $\mathcal{D}$ dataset, i.e., without the direct measurements of $H_0$ or $S_8$. For comparison, we also show the fit of \lcdm\, to the same data and the 68\% and 95\% confidence bands of the direct measurements in gray. One sees clearly that the posterior of WZDR+ is much wider in both $H_0$ and $S_8$ than the one for \lcdm. In addition, the mean of the posterior for $H_0$ is shifted by $1.6$~km/s/Mpc and the one for $S_8$ by 0.01. As a result the posterior  now has significant overlap with the direct measurements. Note also the strongly non-Gaussian tail of the 1D posterior towards smaller values of $S_8$ corresponding to increasing DM-DR interaction $\Gamma_0$, and the much broader and also somewhat non-Gaussian 1D posterior towards larger values of $H_0$ corresponding to increased dark radiation fluid $N_{IR}$. The correlations of the WZDR+ parameters $N_{IR}$ and $\Gamma_0$ with $H_0$ and $S_8$ are clearly visible in Fig.~\ref{fig:wzdrPlus_D_triangle}.

Given the broadening and shift towards larger $H_0$ and smaller $S_8$ we expect to find that the predictions for the values of these parameters in WZDR+ are in less tension with the direct measurements than in $\LCDM$. 

\begin{figure}[t]
	\centering
	\includegraphics[width=0.45\textwidth]{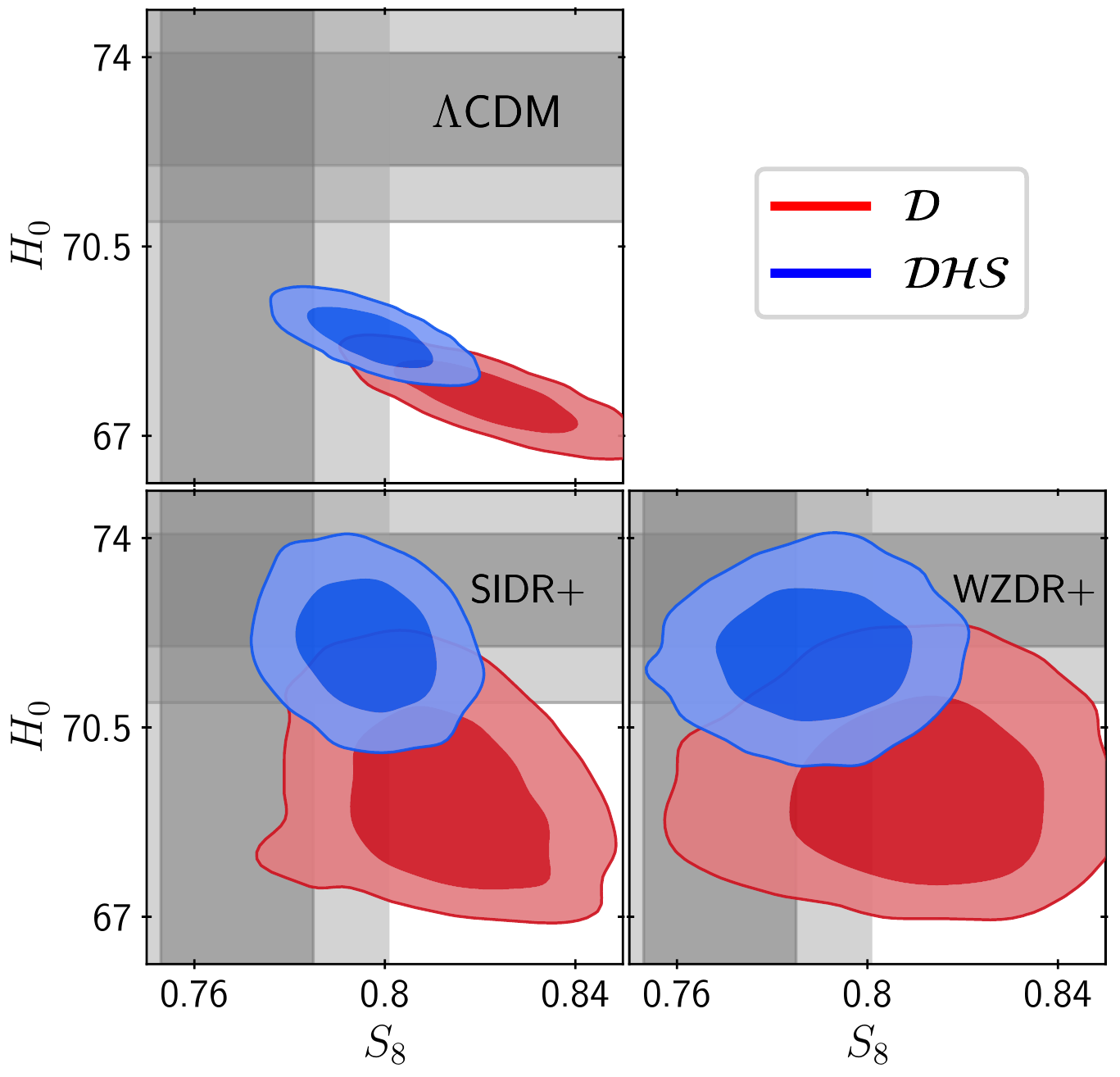} 
	\caption{Posterior distributions of  $\Lambda$CDM ({top left}), SIDR+ ({bottom left}), and WZDR+ ({bottom right}) fitted to $\mathcal{D}$ vs. $\mathcal{DHS}$ datasets. The gray bands show the one and two sigma regions of $H_0$ from the SH0ES Collaboration, and of our combination $S_8^{\rm direct}$.}
	\label{fig:wzdrPlus vs sidrPlus vs LCDM}
\end{figure}

One way to quantify the tension (or lack thereof) between two datasets in a given model is to perform a combined fit to the two datasets in question and examine the goodness of fit, the $\chi^2$. Therefore we performed fits of all three models, $\LCDM$, SIDR+ and WZDR+ to the full dataset \dataDHS. Fig.~\ref{fig:wzdrPlus vs sidrPlus vs LCDM} shows the resulting posteriors in the $\{S_8,\,H_0\}$ plane in blue compared with the fit to the base dataset \dataD (red).
One sees that even with the pull from the direct measurements the $\Lambda$CDM posterior remains far from the direct measurements in $H_0$ and to a lesser extent in $S_8$. 
On the contrary, the much broader WZDR+ posterior from the fit to $\mathcal{D}$ overlaps both direct measurements at $1\sigma$ and almost reaches the overlap of both. Thus it is easily pulled to largely overlap with both direct measurements once fit to the full dataset \dataDHS.
The figure shows that SIDR+ can also address both tensions, and based on the figure alone one cannot ascertain a preference for WZDR+ versus SIDR+ or quantify the goodness of fit of either.

Thus, we need to compute and compare the $\chi^2$ values of the various best-fit points to probe if they provide good overall fits to the data. Specifically, we will compute the $Q_{DMAP}$ value which quantifies the tension between the prediction for an observable from a fit in a model to the direct measurement by comparing the $\chi^2$ values of the best-fit points in the fit with and without the direct measurement. For example, to determine the $H_0$ tension in $\LCDM$, we compare the $\chi^2$ of the best-fit point in the fit to the \dataD dataset, $\chi^2_\mathcal{D}$, to the best-fit point of the fit to the \dataDH dataset, $\chi^2_\mathcal{DH}$. The $Q_{DMAP}$ value%
\footnote{When the direct measurement consists of multiple measurements as in the case of $S_8$ one must also subtract the $\chi^2_\mathcal{S}$ due to the tension between the different direct measurements, and the $Q_{DMAP}$ formula becomes more symmetric $\left(\chi^2_\mathcal{DS}-\chi^2_\mathcal{D}-\chi^2_\mathcal{S}\right)^{1/2}$. Because of the excellent agreement between the KiDS and DES measurements this correction is numerically insignificant $\chi^2_\mathcal{S}=0.08$. For Gaussian posteriors the $Q_{DMAP}$ value agree with the Gaussian tension.}
in units of $\sigma$ is then $\left(\chi^2_\mathcal{DH}-\chi^2_\mathcal{D}\right)^{1/2}$. Assuming Gaussian distributed errors the expectation for the $Q_{DMAP}$ value in a model which perfectly describes the data  is $1\sigma$.

Table~\ref{tb:minitab_QDMAP} shows the results of three different tests, the $Q_{DMAP}$ for the datasets \dataDH\!, \dataDS\!, and \dataDHS\!, all compared to \dataD. Beginning with the first row, we see that within $\LCDM$ the prediction for $S_8$ from the fit to \dataD is in moderate $2.6\sigma$ tension with the direct measurements. Much more significant at $5.6\sigma$ is the Hubble tension, the tension with the direct measurement from SH0ES.\footnote{This value is even larger than the $5\sigma$ tension obtained for $H_0$ in~\cite{Riess:2021jrx}. This is because we quantify the tension with the supernova magnitude $M_b$ instead of $H_0$ which avoids the model dependence  included in the systematic uncertainties of $H_0$ from~\cite{Riess:2021jrx}.} The tension with both direct measurements combined is $5.8\sigma$, and clearly $\LCDM$ cannot explain the combined $H_0/S_8$ tension.\footnote{This is slightly smaller than the combination (in quadrature) of the $Q_{DMAP}$ tensions for \dataDH and \dataDS. This is due to the correlation of $S_8$ and $H_0$ visible in the $\LCDM$ panel of Fig.~\ref{fig:wzdrPlus vs sidrPlus vs LCDM}.}

\begin{table}[h!]
	\setlength\extrarowheight{3pt}
	\centering
	\begin{tabular}{|c || c | c | c |} 
		\hline
		Model/$\QDMAP$& \dataDH & \dataDS & \dataDHS \\
		\hline \hline
		$\text{$\Lambda$CDM}$& $5.57\,\sigma$  & $2.61\,\sigma$   & $5.80\,\sigma$\\  \hline
		SIDR+ & $3.18\,\sigma$ &  $2.79\,\sigma$ & $3.62\,\sigma$\\ \hline
		WZDR+ & $2.45\,\sigma$ & $2.06\,\sigma$  & $3.20\,\sigma$ \\
		\hline
	\end{tabular}
	\caption{$Q_{DMAP}$ tensions}
	\label{tb:minitab_QDMAP}
\end{table}

SIDR+, in contrast, makes a significant improvement in addressing the Hubble tension (which is not surprising, given the extra radiation). It does not help the $S_8$ tension, however, as shown by the lack of improvement in the ${\cal DS}$ dataset. The failure of SIDR+ to significantly reduce $S_8$ can be understood from Fig.~\ref{fig:Pk effects} which shows that the suppression of the MPS starts too early at $k\sim 0.01 h/\mpc$, which generates a tension with CMB data.\footnote{The SIDR+ \dataDS posterior is bimodal, one mode has a minimal amount of extra radiation $N{\rm fluid} \sim 0.0007$ while the other has  $N{\rm fluid} \sim 0.07$ with a local minimum of $\chi^2$. Here we focus on the region of parameter space with $N_{\rm fluid}>0.01$ since we are interested in the models' potential for solving both tensions.} 

The improvement in the ${\cal DHS}$ $Q_{DMAP}$ for SIDR+ is almost entirely driven by the pull of the $\mathcal{H}$ dataset and the reduction of the Hubble tension. WZDR+, on the other hand, does better in all regards, improving ${\cal DS}$ to a two sigma anomaly, and reducing ${\cal DH}$ to below three sigma. The overall cosmological tension from both Hubble and $S_8$ tensions is reduced to about three sigma. If one looks at the breakdown of $\chi^2$ values in Appendix~\ref{app: triangle plots and tables}, one sees that in fitting to \dataDHS, WZDR+ is gaining improvements from all portions of the dataset.

Another test is $\Delta AIC$, the Akaike information criterium, which is defined as the difference of the best-fit $\chi^2$ to all the data $\mathcal{D}HS$ 
between a given model and the reference $\Lambda$CDM, with a $\chi^2$ penalty of $+2$ for each new model parameter beyond $\Lambda$CDM:
\begin{align}
	\Delta AIC & = \chi^2 - \chi^2_{\Lambda \rm CDM} + 2 \times (\text{new parameters})~.
\end{align}

The results of this relatively straightforward test are shown in Table~\ref{tb:minitab_AIC}. We see immediately that the inclusion of just a single parameter (the DM interaction strength versus WZDR, and the mass threshold versus SIDR+) improves the $\chi^2$ by more than 5, improving the $\Delta AIC$ in each case by better than 3. This is a weak preference, to be clear, but is also reflecting the simple fact that the $S_8$ data are not (yet) strong enough to make this more than a tension.
\begin{table}[h!]
	\setlength\extrarowheight{3pt}
	\centering
	\begin{tabular}{|c | c | c | c |} 
		\hline
		& WZDR & SIDR+ & WZDR+ \\
		\hline 
		parameters  & $2$  & $2$  & $3$\\  \hline
		$\Delta \chi^2 $ & $-20.52$  & $-19.99$  & $-25.78$\\  \hline
		$\Delta AIC $ & $-16.52$  & $-15.99$  & $-19.78$\\  \hline
	\end{tabular}
	\caption{$\chi^2$ differences and $\Delta AIC$ of WZDR, SIDR+, and WZDR+ relative to $\LCDM$}
	\label{tb:minitab_AIC}
\end{table}

Finally, there is the Gaussian tension (GT) test. This is not an ideal test in this case because the posteriors for $S_8$ and the supernova magnitude $M_b$ are not Gaussian in SIDR+ and WZDR+, but we nonetheless include the GT for completeness \footnote{The Gaussian tension between two measurements with their $1\sigma$ errors $x_i\pm \delta x_i$ is defined as $GT=|x_1-x_2|/\sqrt{\delta x_1^2 + \delta x_2^2}$.}. The posteriors for SIDR+ and WZDR+ overlap the direct measurements of $S_8$ and $M_b$ at the $\sim ~ 2-3\, \sigma$ level, therefore we use one half of the $2\sigma$ intervals characterizing the 1-d posteriors (these are produced in the 
`.h\_info' files in the analysis output of MontePython). This gives a slightly better approximation to the true tension in the models than simply using the $1\sigma$ intervals. 

Table~\ref{tb:minitab_GT} shows that the predicted value for $M_b$ from the fit to $\mathcal{D}$ in $\LCDM$ has a GT of about $5.5\sigma$ to the direct measurement of $M_b$ from SH$_0$ES. In WZDR+ (and SIDR+) this tension is reduced to $2.6\sigma$ ($2.7\sigma$) due to the interacting radiation with (and without) a step. Similarly, Table~\ref{tb:minitab_GT} shows a reduction of the GT in $S_8$ (here we compare to the combined KiDS+DES measurement) from $2.6\sigma$ in $\LCDM$ to $1.7\sigma$ in WZDR+ and $2.0\sigma$ in SIDR+. This reduction in $S_8$ is due to the DMDR interaction.\footnote{Using the naive $1\sigma$ intervals, the GT for $M_b$ is $5.5/3.2/3.0$, and for $S_8$ it is $2.6/2.0/1.8$ in order $\LCDM$/SIDR+/WZDR+.}

\begin{table}[h!]
	\setlength\extrarowheight{3pt}
	\centering
	\begin{tabular}{|c || c | c |} 
		\hline
		Model/GT & $~M_b~$ & $~S_8~$  \\
		\hline \hline
		$\text{$\Lambda$CDM}$&  $5.5\,\sigma$ & $2.6\,\sigma$   \\  \hline
		SIDR+ &   $2.7\,\sigma$ & $2.0\,\sigma$ \\ \hline
		WZDR+ & $2.6\,\sigma$ & $1.7\,\sigma$ \\
		\hline
	\end{tabular}
	\caption{Gaussian tensions.}
	\label{tb:minitab_GT}
\end{table}

The success we see above (in reducing the combined tensions in the data), still leaves the question whether the new ingredient -- namely the transition scale -- is working simultaneously to alleviate both tensions, whether they pull in different directions, or whether the improvements are really independent of each other. 
As was discussed in previous sections the WZDR+ model is sensitive to the mass threshold at $z_t$ through two independent physical processes (a) via the $\ell$ dependence of the phase shift of the CMB due to the change in $N_{\rm eff}$ when $T_d$ drops below $m_\phi$, and (b) through the suppression of the matter power spectrum from the coupling between the dark matter and the dark radiation fluid due to scattering at temperatures above $m_\phi$. In our previous paper we found that the CMB preferred $\log_{10} (z_t) \sim 4.3$, and it is interesting to see if this value of $z_t$ is also preferred by the \dataDS dataset; the CMB power spectrum, the CMB lensing potential and the matter power spectrum at distance scales of order $k_8\sim$ h/(8Mpc) which are all sensitive to the shutoff redshift of the interaction.

\begin{figure}[t]
	\includegraphics[width=0.45\textwidth]{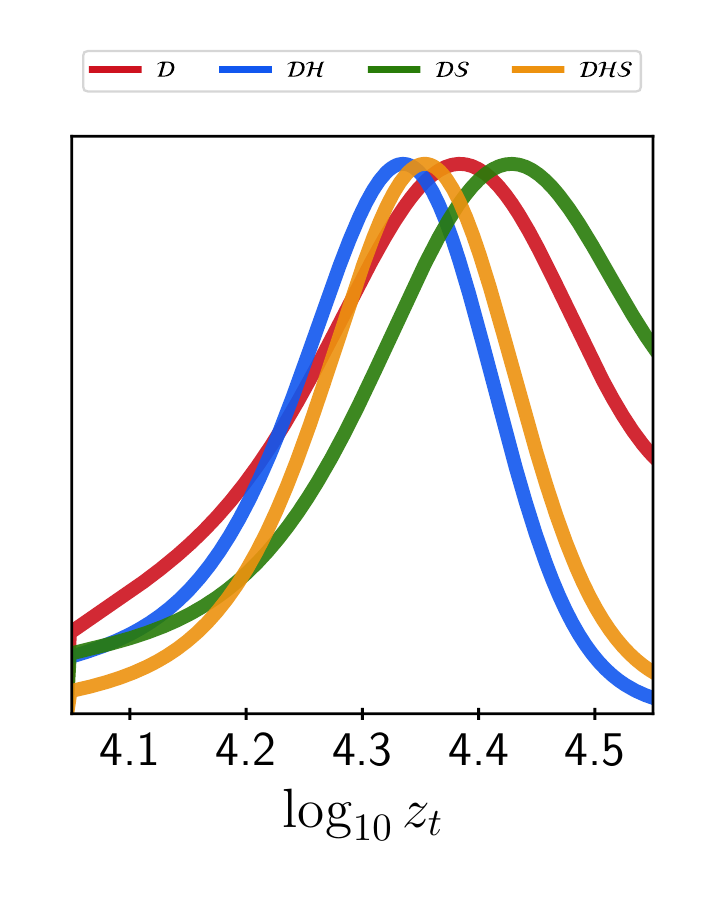}
	\caption{The 1D posteriors of the transition time $z_t$, fitting WZDR+ to four different datasets.}
	\label{fig:zt}
\end{figure}

To answer this question we compare the WZDR+ mean-values and best-fit points for the four datasets $\mathcal{D}, \mathcal{DH}, \mathcal{DS},$ and $\mathcal{DHS}$ and check for consistency. 
\begin{table}[h!]
	\setlength\extrarowheight{3pt}
	\centering
	\begin{tabular}{|c | c | c | c |c |} 
		\hline
		$\log_{10} z_t$ & $\mathcal{D}$ & $\mathcal{D}H$  & $\mathcal{D}S$ & $\mathcal{D}HS$ \\
		\hline 
		Mean & $4.35^{+0.17}_{-0.12}$  & $4.29^{+0.12}_{-0.08}$  & $4.38^{+0.17}_{-0.09}$ & $4.32^{+0.11}_{-0.09}$ \\  \hline
		Best-fit & $4.33$  & $4.26$  & $4.38$ & $4.32$ \\  \hline
	\end{tabular}
	\caption{Mean and $\pm 1\,\sigma$, and best-fit values of $log10 z_t$ in WZDR+ fitted to the four different datasets.}
	\label{tb:minitab_log10_zt}
\end{table}
We find the remarkable result that the value of $z_t$ preferred by dataset $\mathcal{D}$ alone (which is dominated by the CMB) is the same to within half a sigma to the preferred value for $\mathcal{DH}$, $\mathcal{DS}$ and $\mathcal{DHS}$ even though $H_0$ and $S_8$ shift by 2.8$\sigma$ and 1.4$\sigma$, respectively.
We find this coincidence interesting and consider it potential evidence for the existence of a new scale $T_d\sim10$ eV in Cosmology corresponding to redshifts of order $z_t\sim3\times 10^5$. This coincidence can also been seen in the 1D posteriors of the variable $\log_{10} (z_t)$ for the four different datasets shown in Fig.~\ref{fig:zt} and more concretely in the values of the best-fit values of $z_t$ seen in Table~\ref{tb:minitab_log10_zt}.  Further information regarding the posterior distributions and best-fit points is presented in the Figures and Tables of Appendix~\ref{app: triangle plots and tables}.
%

\section{Discussion and Outlook}\label{sec:conclusions}
In the energy range $\Lambda_{QCD} < E< \tev$, the Standard Model has seven mass thresholds including a phase transition. Below $\Lambda_{QCD}$, the Standard Model has myriad mass thresholds from the resonances of quarks, but additionally the muon and electron masses, and at least two neutrino masses. It is somewhat striking, then, that it is commonly assumed that models of dark sectors exhibit no meaningful mass thresholds when all the physics we have ever seen is full of them.

In this vein, we have considered the effect of a single mass threshold in a dark fluid which is  gently coupled to the dark matter (the WZDR+ model). The mass threshold has been previously shown to allow for enhanced $H_0$ with better overall fit when compared to $\Lambda$CDM or to a fluid with no mass threshold (WZDR). In particular, the Hubble tension seems to suggest a mass threshold near $z_t \sim 20,000$, when the sound horizon is approximately $10h^{-1} \mpc$.

Even without knowing this, it would be natural to consider extensions to the WZDR model where the dark radiation is coupled to the dark matter. However, noting that the threshold occurs at a location which is precisely in the range of {\it another} known anomaly, namely the $S_8$ tension, suggests that the two anomalies -- the $H_0$ tension and the $S_8$ tension, may have a common origin.

As the dark radiation passes through the mass threshold, the gentle coupling to the dark sector turns off rapidly. This can occur either because the particle with which the DM is interacting has become exponentially suppressed, or because the mediator mass is suddenly relevant, and the gentle interaction turns off. This has the natural effect of producing a CDM cosmology on large scales, and non-CDM on small scales, with the transition occurring at a scale which is singled out by the $H_0$ tension to be $\sim 10\, \mpc$. We have seen that this scenario naturally produces a suppressed value of $S_8$, consistent with the directly observed value. 

It is quite striking that all of the different combinations of datasets (${\cal D, DS, DH, DHS}$), all point to the same value of $z_t$ (although the preference within ${\cal D}$ alone is quite small). But this is a nontrivial consistency check, without which this overall setup would be unable to reconcile these observations.

Our efforts here are the simplest extension of WZDR. One could consider multi-component dark matter where only a portion couples to the dark sector, but with enhanced interaction~\cite{Chacko:2016kgg,Buen-Abad:2017gxg}. In such a ``fractional WZDR+'' setup, one would expect that a similar phenomenology to the above would be found, but constraining a product of interaction strength and interacting dark matter fraction, leaving a large degeneracy. In the tightly coupled limit, the interacting dark matter fraction would acoustically oscillate, rather than feel a slight friction during infall. In this limit, one would similarly expect a good fit, but trading a precise value of $\Gamma_0$ for a precise value of the interacting fraction $f_\chi$ to fit the value of $S_8$. We leave the details to future work.

Beyond this, one could also imagine multiple mass thresholds, couplings to neutrinos and more. What is clear, however, is that the data, in their present form, provide sensitivity to the presence of a mass threshold in the dark sector. As data improve - both from CMB datasets of ACT, Simons Observatory and CMB-S4, as well as from LSS measurements KiDS, DES, HSC, and future galaxy surveys with Rubin, Roman, and UNIONS -- it will become clear both whether a dark mass threshold is truly preferred by the data, and what sort of dark sector dynamics we are being pointed to.

\vskip 0.3in 

{\bf Note added:} As this work was being completed, \cite{Schoneberg:2022grr} appeared, which 
also studies DM interacting with a WZDR fluid and includes fits to additional data. Importantly, \cite{Schoneberg:2022grr} do not consider the $z_t$-dependent turnoff of the DM-DR interaction which we studied in this paper.

\section*{Acknowledgements}
We thank Asher Berlin for collaboration and insightful discussions about phase shifts at the beginning of this work. The work of D. A., M. J., M. S. and E. N. S. is supported by the U.S. Department of Energy (DOE) under Award No. DE-SC0015845. N.W. is supported by NSF under Award No. PHY-2210498, by the BSF under Grant No. 2018140, and by the Simons Foundation. Our MCMC runs were performed on the Shared Computing Cluster, which is administered by Boston University’s Research Computing Services.

\bibliographystyle{utphys}
\bibliography{WZDRplus}
\appendix

\section{Models and Acronyms}\label{app: acronyms}
\begin{table}[h!]
	\setlength\extrarowheight{3pt}
	\centering
	\begin{tabular}{l c p{0.8\columnwidth} } 
		$\Lambda$CDM & - & The concordance model. \\ 
		SIDR & - & $\Lambda$CDM + self-interacting dark radiation fluid.\\
		SIDR+ & - & SIDR + dark radiation interacting with the dark matter. \\
		WZDR & - & A model in which the energy density of the dark radiation fluid increases during the time of the CMB~\cite{Aloni:2021eaq}.\\
		WZDR+ & - & WZDR + dark radiation interacting with the dark matter. \\
	\end{tabular}
	\label{tb:models}
\end{table}

\section{Momentum transfer rate}\label{app: momentum transfer rate}
The perturbation equations for the dark matter and the WZDR fluid are sensitive to the momentum transfer rate between the two fluids. 
This rate is defined as~$\dot{\vec{P}}_{\chi} = -a \Gamma \vec{P}_{\chi}$, the change in momentum of a DM particle $P_\chi$ due to friction it experiences while moving through the WZDR fluid of temperature $T$. The thermally averaged rate is given by
\begin{widetext}
	------	\begin{align}\label{eq: general momentum transfer rate}
		\dot{\vec{P}} & = \frac{a}{2 E_P} \int\frac{d^3 k}{(2\pi)^3 \, 2 E_k} f(k;T) \int\frac{d^3 k'}{(2\pi)^3 \, 2 E_k'} \frac{d^3 P'}{(2\pi)^3 \, 2 E_P'}
		(2\pi)^4 \delta^{(4)}(P + k - P' -k') \left|\mathcal{M}\right|^2\left(\vec{P}\,' -  \vec{P}\right)~,
	\end{align}
\end{widetext}
where $P,\,P'$ stands for the incoming and outgoing DM momentum, $k,\,k'$ stands for a WZDR momentum, $f(k;T)$ is the thermal-distribution function for the incoming scatterers from the thermal bath, and we neglect the stimulated emission/Pauli blocking term for final-state particles.

As discussed in the text the $\phi-\chi$ scattering is suppressed by the massive DM propagator. Here we consider only the $\psi-\chi$ scattering mediated by $t$-channel $\phi$ exchange. The matrix element relevant for this process has the following dependence on the kinematical variables
\begin{align}
	\left|\mathcal{M}\right|^2 & = \frac{g_{\chi\phi}^2 g_{\psi\phi}^2}{4} \frac{t(t - 4M^2)}{(t - m_\phi^2)^2}
	~,
\end{align}
where the subscript of the coupling constants indicate the particles involved in the interaction.  

Plugging this matrix element into Eq.~\eqref{eq: general momentum transfer rate}, and after some tedious algebra one finds
\begin{align}
	\Gamma & = \tilde{\alpha}^2\frac{T^2}{M} \int_{0}^{\infty} d\tilde{k} \,  \tilde{k} ^2   e^{-\tilde{k} }\int
	\frac{ dc_\theta \, (1-c_\theta)^2}{\left[2(1-c_\theta) + \frac{x^2}{\tilde{k}^2}\right]^2}~,
\end{align}
where $x = m_\phi / T$, and $\tilde{\alpha}$ is an effective average coupling constant normalized such that $\Gamma=\tilde{\alpha}^2\frac{T^2}{M}$ in the case of $x= 0$ . The integrals above can be evaluated analytically in terms of an awful expression with various special functions, but in all regions of interest  it is approximated to a few percent precision by
\begin{align*}
	\left(\frac{1}{1 - 0.05 \sqrt{x} + 0.131 x }\right)^4 ~,
\end{align*}
where the coefficients have been tuned to approximate the exact result.
Finally we parametrize the momentum transfer rate as
\begin{align}
	\Gamma(x) & = \Gamma_0\frac{(1+z_t)^2}{x^2} \left(\frac{1}{1 - 0.05 \sqrt{x} + 0.131 x }\right)^4 ~,
\end{align}
where $\Gamma_0$ is the momentum transfer rate extrapolated to today in a theory in which there is no step (i.e. where $m_\phi=0$). 

With this definition the interaction rate in the UV (i.e. $T\gg m_\phi$) is
$\Gamma\simeq\Gamma_0 (T_{UV}/T_0)^2=\Gamma_0 (1+z)^2 (7/15)^{2/3}$ so that the scattering rate in the UV is smaller than the corresponding rate in SIDR+ by $(7/15)^{2/3}\sim 0.6$.

For completeness, the momentum transfer rate enters into the dipole equations of the interacting  DM and the Wess-Zumino fluid as
\begin{align}
	\dot{\theta}_{dm} & = -\mathcal{H}\theta_{dm} + k^2 \psi + a\Gamma \left(\theta_{wz} - \theta_{dm}\right)~, \\
	\dot{\theta}_{wz} & = k^2 \left(\frac{\delta_{wz}}{4} + \psi\right) - a\Gamma R \left(\theta_{wz} - \theta_{dm}\right)~, 
\end{align}
where $R \equiv 3\rho_{dm}/ 4\rho_{wz}$.

\section{Sensitivity to nonlinear effects}\label{app: halofit}
Since our fits include the $S_8$ observable which is sensitive to the MPS at (weakly) nonlinear scales, we briefly investigate the potential for bias due to poorly modeled nonlinear effects. 

The concern is that extraction of $S_8$ from the experimental data requires modeling of the nonlinear effects in the growth of structure. The models which are used, {\it HMcode}~\cite{Mead:2015yca} and {\it halofit}~\cite{Smith:2002dz,Takahashi:2012em}, have been tuned to correctly reproduce the results of $N$-body simulations for $\Lambda$CDM but may not be reliable for physics beyond $\Lambda$CDM and therefore the experimentally quoted values for $S_8$ may be biased when applied to other models. 

\begin{figure}[t]
 	\includegraphics[width=0.45\textwidth]{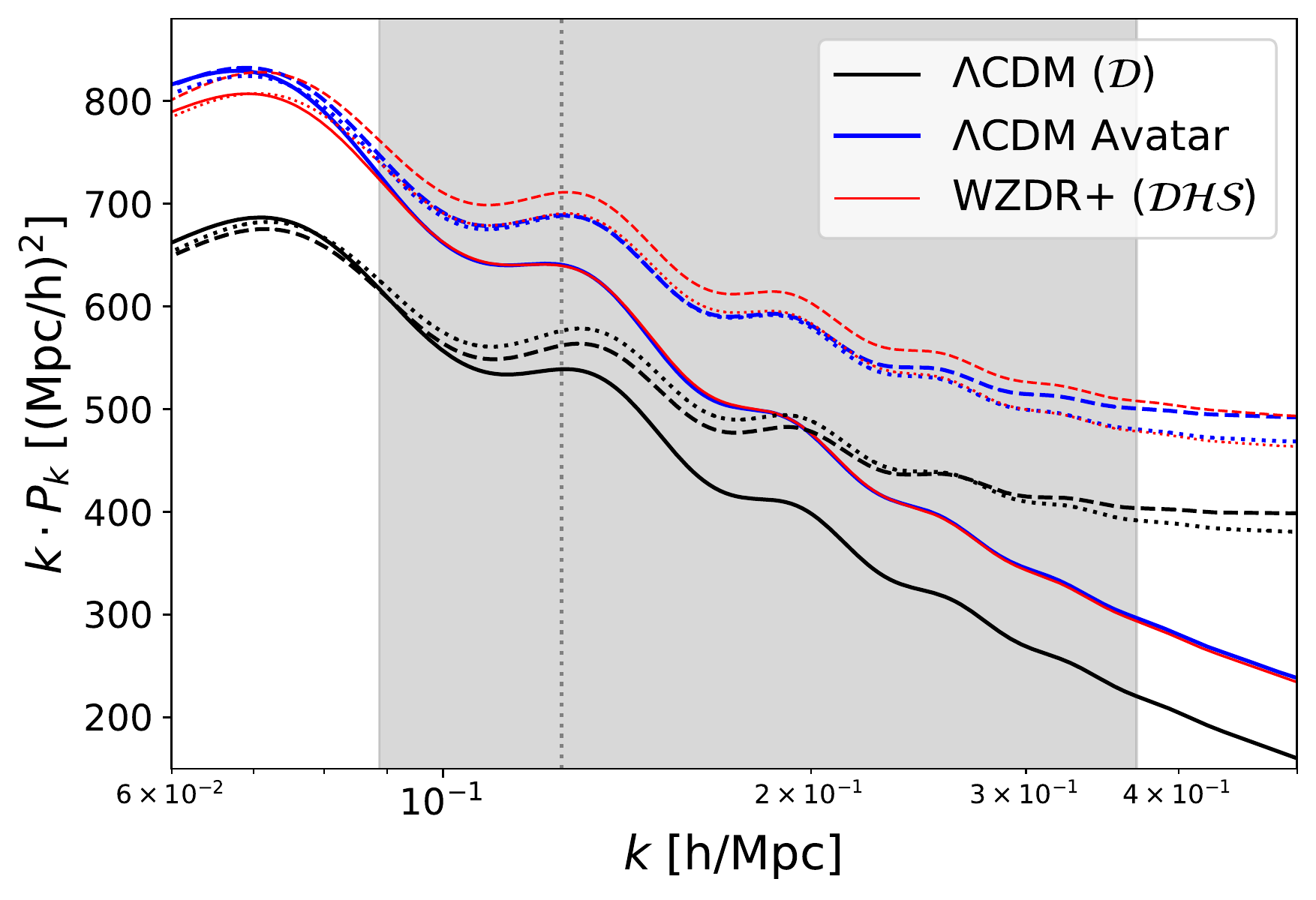}
 	\caption{Matter power spectra of the best-fit point of WZDR+ for $\mathcal{DHS}$ dataset (red), compared with its $\Lambda$CDM Avatar (blue), 
 		and the best-fit point of  $\Lambda$CDM for the $\mathcal{D}$ dataset (black). The linear MPS are shown as solid lines, the nonlinear MPS calculated with {\it halofit} as dashed lines, and the nonlinear MPS calculated with {\it HMcode} as dotted lines. For WZDR+ and the Avatar $k\cdot P(k)$ is shifted vertically by 100 $(Mpc/h)^2$ for clarity.}
 	\label{fig:MPS_linear_vs_nonlinear}
 \end{figure}

To investigate this issue for WZDR+, we note that the growth of structure at scales relevant to $S_8$ ($0.05\lsim k \lsim 0.5$) incurs significant nonlinear corrections at late times, near $z=0$. This is well after the WZDR+ interactions have turned off at $z_t\sim 2 \times 10^4$ and when DR energy densities are negligible. Therefore we expect that late time growth in WZDR+ should be well-approximated by growth in a $\Lambda$CDM ``Avatar'' model which has a primordial matter power spectrum chosen so that the linear MP spectra of WZDR+ and the Avatar match for the $k$-range of interest for $S_8$.
In Fig.~\ref{fig:MPS_linear_vs_nonlinear}, we show that the linear MPS of WZDR+ fit to the $\mathcal{DHS}$ dataset (solid red) is well-matched by a $\Lambda$CDM Avatar (solid blue) with slope $n_s=0.87$ and amplitude $\ln 10^{10} A_s=3.09$ (and all other cosmological parameter values equal to the best-fit $\Lambda$CDM in the $\mathcal{DHS}$ dataset). The Avatar's values for $n_s$ and $A_s$ are clearly outside the range preferred by fits to the Planck CMB data, but they are within the range of $\Lambda$CDM parameters studied in the $S_8$ analyses of the KiDS Collaboration~\cite{Heymans:2020gsg} and DES-Y3~\cite{DES:2021wwk}. Thus if the nonlinear corrections are modeled reliably by the experiments for $\Lambda$CDM we can also expect them to be reliable for WZDR+.

To support this expectation, in Fig.~\ref{fig:MPS_linear_vs_nonlinear} we also plot the nonlinear matter power spectra for WZDR+ and the Avatar computed with CLASS and choosing either {\it HMcode} (dotted) or {\it halofit} (dashed). For {\it HMcode} the nonlinear MP spectra of WZDR+ and its $\Lambda$CDM Avatar are almost identical. For {\it halofit} we find differences which are only slightly larger than the differences between {\it HMcode} and {\it halofit} in ordinary $\Lambda$CDM. For comparison, we also show the MPS of the best-fit point of $\Lambda$CDM for the $\mathcal{D}$ dataset (black) for which {\it HMcode} and {\it halofit} reproduce $N$-body simulations.

In conclusion, we see no indication that the nonlinear growth of structure in the parameter range of interest for $S_8$ is modeled incorrectly by {\it HMcode} or {\it halofit} in WZDR+.
However, given that $S_8$ does have support on nonlinear scales, it would be interesting to perform a full-shape analysis using the effective theory of large-scale structure in future work.

\newpage
\section{Triangle Plots and Parameter Values}\label{app: triangle plots and tables}
\begin{figure*}[!htbp]
	\centering
	\includegraphics[width=1.\textwidth]{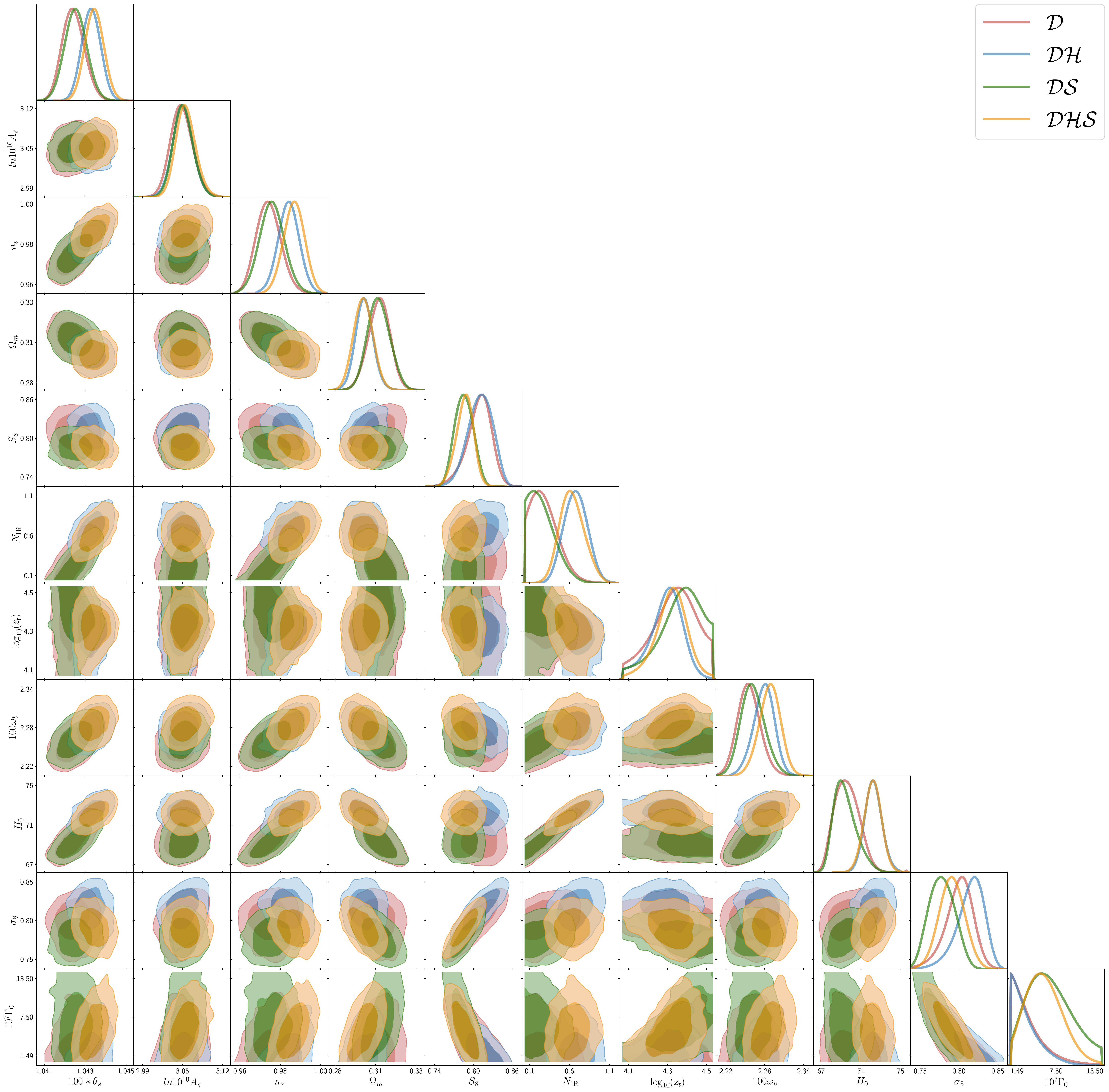}
	\caption{A comparison of the posteriors of WZDR+ for the four different datasets. The dark and light shaded regions correspond to 68\% and 95\% C.L., respectively. We see that the preference for the location of the transition ($\zt$) is fairly consistent across the datasets possibly signaling new physics at this scale.  Additionally the fit to $\mathcal{DHS}$ in comparison with the fits to $\mathcal{DH}$ and $\mathcal{DS}$ show that WZDR+ is capable of simultaneously alleviating the $H_0$ and  $S_8$ tensions. }
	\label{fig:wzdr-4param-triangle}
\end{figure*}

\begin{figure*}[!htbp]
	\centering
	\includegraphics[width=1.\textwidth]{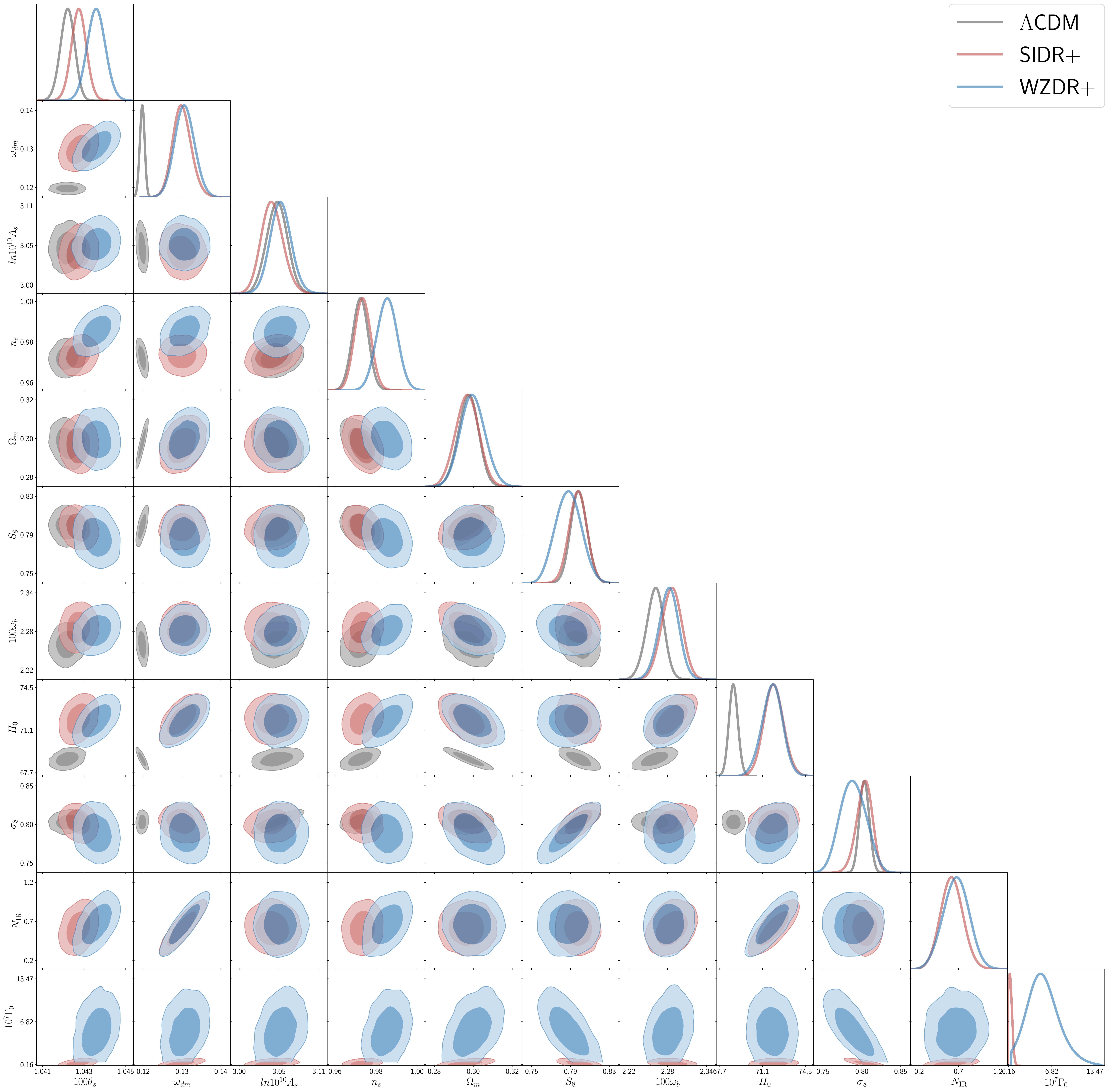}
	\caption{A comparison of the posteriors $\LCDM$, SIDR+, and WZDR+ fitting to $\mathcal{DHS}$. The dark and light shaded regions correspond to 68\% and 95\% C.L., respectively. From this comparison we see the importance of the step in simultaneously alleviating the $H_0$ and $S_8$ tensions. Although SIDR+ allows for larger $H_0$ values, it does no better than $\LCDM$ in resolving the $S_8$ tension.}
	\label{fig:3model-triangle}
\end{figure*}

In this appendix, we provide supplemental information to the results of our analysis of Sec. IV. 2D posterior distributions are presented in Figs.~\ref{fig:wzdr-4param-triangle} and~\ref{fig:3model-triangle}, while best-fit values and means are provided in Tables~\ref{tb:mean_and_bf_D_DHS}-\ref{tb:bf_for_qdmap}.

\begin{table*}[h!]{ }
	\centering
	\begin{tabular}{|c | c | c | c | c | c | c|} 
		\hline
		& \multicolumn{2}{c|}{$\Lambda{\rm CDM}$} &
		\multicolumn{2}{c|}{SIDR+} &
		\multicolumn{2}{c|}{WZDR+} \\
		\hline
		& $\mathcal{D}$ & $\mathcal{D}$HS &$\mathcal{D}$ &  $\mathcal{D}$HS &$\mathcal{D}$ & $\mathcal{D}$HS  \\
		\hline &&&&&&\\[-8pt]
		100$\theta_s$ & $1.04193^{+0.00028}_{-0.00029}$ & $1.04216^{+0.00030}_{-0.00029}$  & $1.04220^{+0.00031}_{-0.00034}$ & $1.04261^{+0.00030}_{-0.00032}$ & $1.04260^{+0.00044}_{-0.00047}$& $1.04347^{+0.00040}_{-0.00040}$  \\  [3pt]
		$\Omega_b h^2$ & $0.02240^{+0.00014}_{-0.00014} $ & $0.02263^{+0.00013}_{-0.00014}$  & $0.02254^{+0.00016}_{-0.00020}$ & $0.02288^{+0.00015}_{-0.00015}$ &  $ 0.02256^{+0.00016}_{-0.00017}$ & $0.02287^{+0.00016}_{-0.00016}$ \\ [3pt]
		$ \Omega_{\rm dm}h^2$ & $ 0.11911^{+0.00097}_{-0.00098}$ & $0.11678^{+0.0079}_{-0.0082}$ & $0.1237^{+0.0024}_{-0.0038} $ & $0.1295^{+0.0028}_{-0.0032}$ & $0.1244^{+0.0025}_{-0.0038}$ & $0.1305^{+0.0031}_{-0.0034} $ \\ [3pt]
		$\ln 10^{10} A_s$  & $3.045^{+0.015}_{-0.017} $ & $3.050^{+0.015}_{-0.015} $ & $3.043^{+0.017}_{-0.018}$ & $3.043^{+0.015}_{-0.017} $ & $3.050^{+0.015}_{-0.018}$ & $3.057^{+0.014}_{-0.017}$ \\ [3pt]
		$n_s$ &  $0.9662^{+0.0039}_{-0.0038} $ & $0.9717^{+0.0038}_{-0.0035}$& $0.9681^{+0.0040}_{-0.0038} $ &$0.9727^{+0.0037}_{-0.0037} $ & $0.9742^{+0.0052}_{-0.0057}$ &$0.9861^{+0.0051}_{-0.0048}$\\ [3pt]
		$\tau_{\rm reio}$ & $0.0558^{+0.0075}_{-0.0080} $ & $0.0602^{+0.0076}_{-0.0077}$ & $0.0563^{+0.0077}_{-0.0084} $ & $0.0601^{+0.0072}_{-0.0083} $ & $0.0568^{+0.0073}_{-0.0082}$ & $0.0587^{+0.0069}_{-0.0082}$ \\ [3pt]
		$\nir $  &-&-& $0.252^{+0.068}_{-0.239}$& $0.63^{+0.14}_{-0.14} $ & $0.29^{+0.10}_{-0.24}$ &  $0.67^{+0.15}_{-0.16}$ \\ [3pt]			
		$10^7\Gamma_0 $ [1/Mpc]  &- &-& $0.254^{+0.060}_{-0.254} $ & $0.43^{+0.16}_{-0.38} $ & $2.95^{+0.63}_{-2.95}$ & $5.7^{+2.4}_{-3.4}$  \\ [3pt]
		$\log_{10} (\zt)$  &- &- &- &-& $4.35^{+0.17}_{-0.11}$ & $4.322^{+0.106}_{-0.088}$ \\ [3pt]
		$\sigma 8$ & $0.8085^{+0.0070}_{-0.0073}$& $0.8031^{+0.0058}_{-0.0060}$ & $0.8032^{+0.0132}_{-0.0096} $ & $0.8029^{+0.0110}_{-0.0090}$ & $0.800^{+0.020}_{-0.013}$ & $0.791^{+0.016}_{-0.016}$  \\ [3pt]
		$\Omega_m$ & $0.3101^{+0.0058}_{-0.0060}$& $0.2956^{+0.0046}_{-0.0048}$ & $0.3068^{+0.0067}_{-0.0068} $ &  $0.2950^{+0.0051}_{-0.0054} $ &  $0.3078^{+0.0066}_{-0.0065}$ &$0.2983^{+0.0057}_{-0.0059}$  \\ \hline  &&&&&&\\[-8pt]		
		$M_b $ & ${-19.417}^{+0.013}_{-0.012}$ & ${-19.388}^{+0.010}_{-0.011} $ & ${-19.371}^{+0.026}_{-0.041}$ & ${-19.290}^{+0.023}_{-0.024}$ & ${-19.368}^{+0.026}_{-0.038}$ & ${-19.294}^{+0.023}_{-0.024}$ \\ [3pt]
		$H_0$ [km/s/Mpc] & $67.71^{+0.45}_{-0.44} $ & $68.83^{+0.38}_{-0.37}$ & $69.22^{+0.88}_{-1.37}$ & $72.03^{+0.80}_{-0.81} $ & $69.29^{+0.93}_{-1.21}$ & $71.86^{+0.79}_{-0.83}$  \\ [3pt]
		$S_8$ & $0.822 ^{+0.012}_{-0.013} $ & $0.7972 ^{+0.0087}_{-0.0088}$& $0.812^{+0.015}_{-0.014} $ &  $0.7962 ^{+0.0097}_{-0.0094} $ & $0.810^{+0.020}_{-0.016}$ & $0.7889^{+0.0014}_{-0.0013}$ \\  [3pt] \hline
		
	\end{tabular}
	\caption{Mean and $\pm 1 \sigma$ values for a fit to dataset $\D$ and $\D H S$.}
	\label{tb:mean_and_bf_D_DHS}
\end{table*}

\begin{table*}[h!]{ }
	\centering
	\begin{tabular}{|c | c | c | c | c | c | c|} 
		\hline
		& \multicolumn{2}{c|}{$\Lambda{\rm CDM}$} &
		\multicolumn{2}{c|}{SIDR+} &
		\multicolumn{2}{c|}{WZDR+} \\
		\hline
		&$\mathcal{D}$ & $\mathcal{D}$HS &$\mathcal{D}$ &  $\mathcal{D}$HS & $\mathcal{D}$ & $\mathcal{D}$HS  \\
		\hline &&&&&&\\[-8pt]
		100$\theta_s$ & 1.04193 &1.04217 & 1.04196& 1.04256& 1.04251  &1.04347 \\  [3pt]
		$\Omega_b h^2$ & 0.02239 &0.02265 & 0.02247&0.02287& 0.02255 & 0.02286  \\ [3pt]
		$ \Omega_{\rm cdm}h^2$ & 0.11925 & 0.11685&0.11997 & 0.12868 & 0.1243 & 0.1307 \\ [3pt]
		$\ln 10^{10} A_s$  & 3.044 & 3.052 &3.044 & 3.039& 3.047 & 3.053 \\ [3pt]
		$n_s$ & 0.9666 & 0.9735&0.9677&0.9721& 0.9735 & 0.9867\\ [3pt]
		$\tau_{\rm reio}$ & 0.0547 & 0.0608 & 0.0560 & 0.05870 & 0.0561 & 0.0574 \\ [3pt]
		$\nir $  &-&-&0.057 & 0.60& 0.30 & 0.67 \\ [3pt]			
		$10^7\Gamma_0 $ [1/Mpc]  & -&-& 0.001 & 0.296 &  0.050 & 4.99 \\ [3pt]
		$\log_{10} (\zt)$  &- &- &- &-& 4.33 &4.32 \\ [3pt]
		$\sigma 8$ & 0.8085 & 0.8049 & 0.8094 & 0.8060 & 0.8203 &0.7937 \\ [3pt]
		$\Omega_m$ & 0.3109 & 0.2959 & 0.3081 & 0.2941 & 0.3057 & 0.2987 \\ [3pt] \hline  &&&&&&\\[-8pt]	
		$M_b $ &${-19.419}$ &${-19.387}$ &${-19.404}$&${-19.292}$& ${-19.364}$ & ${-19.294}$ \\ [3pt]
		$H_0$ [km/s/Mpc] & 67.64 & 68.82 & 68.15 &71.93& 69.46 & 71.84 \\ [3pt]
		$S_8$ & 0.823 & 0.799 & 0.820 & 0.7981 &0.828 & 0.7920  \\  [3pt] \hline \hline &&&&&&\\[-8pt]
		$\chi^2_{\rm CMB}$  & $2767.08$ & $2773.04$ & $2767.20$ & $2774.48$ & $2766.15$ &  $2771.55$  \\ [3pt]
		$\chi^2_{\rm Pantheon}$  & $1025.93$ &  $1025.73$ & $1025.80$ & $1025.74$ & $1025.72$ & $1025.64$ \\ [3pt]
		$\chi^2_{\rm BAO}$ & $5.65$ & $6.14$ & $5.32$ & $6.58$ & $5.14$ &  $5.67$ \\ [3pt]
		$\chi^2_{\rm Pl. \ lensing}$  & $(9.07)$ & $10.37$ & $(9.13)$ & $11.31$  & $(9.30)$ & $10.43$ \\ [3pt]
		$\chi^2_{\rm S8}$  & $(11.59)$ & $3.65$ & $(10.41)$ & $3.36$ & $(13.77)$ & $2.12$  \\ [3pt]
		$\chi^2_{\rm SH0ES}$  & $(37.64)$ & $24.61$ & $(31.21)$ & $2.08$ & $(16.81)$ & $2.34$ \\ [3pt]
		\hline &&&&&&\\[-8pt]
		$\chi^2_{\rm tot}$  & $3798.66$ & $3843.54$  & $3798.33$ & $3823.55$ & $3797.01$ & $3817.76$ \\ [3pt]
		\hline
		
	\end{tabular}
	\caption{Best-fit values for a fit to dataset $\D$ and $\D H S$.}
	\label{tb:bestfit_D_DHS}
\end{table*}

\begin{table*}[h!]{ }
	\centering
	\begin{tabular}{|c | c | c | c | c |} 
		\hline
		&$\mathcal{D}$ & $\mathcal{D}$H & $\mathcal{D}$S & $\mathcal{D}$HS  \\
		\hline &&&&\\[-8pt]
		100$\theta_s$ & $1.04260^{+0.00044}_{-0.00047}$& $1.04334^{+0.00038}_{-0.00039}$ & $1.04268^{+0.00043}_{-0.00052}$ & $1.04347^{+0.00040}_{-0.00040}$  \\  [3pt]
		$\Omega_b h^2$ & $ 0.02256^{+0.00016}_{-0.00017}$ & $0.02278^{+0.00017}_{-0.00015}$ & $0.02263^{+0.00016}_{-0.00019}$ & $0.02287^{+0.00016}_{-0.00016}$ \\ [3pt]
		$ \Omega_{\rm dm}h^2$ & $0.1244^{+0.0025}_{-0.0038}$ & $0.1311^{+0.0030}_{-0.0030}$ & $0.1241^{+0.0025}_{-0.0040}$ & $0.1304^{+0.0031}_{-0.0034} $ \\ [3pt]
		$\ln 10^{10} A_s$ & $3.050^{+0.015}_{-0.018}$ & $3.052^{+0.016}_{-0.017}$ & $3.052^{+0.014}_{-0.017}$ & $3.057^{+0.014}_{-0.017}$ \\ [3pt]
		$n_s$ & $0.9742^{+0.0052}_{-0.0057}$ & $0.9836^{+0.0047}_{-0.0050}$ & $0.9758^{+0.0056}_{-0.0063}$ &$0.9861^{+0.0051}_{-0.0048}$\\ [3pt]
		$\tau_{\rm reio}$ & $0.0568^{+0.0073}_{-0.0082}$ & $0.0578^{+0.0073}_{-0.0085}$ & $0.0570^{+0.0072}_{-0.0080}$ & $0.0587^{+0.0069}_{-0.0082}$ \\ [3pt]
		$\nir $ & $0.29^{+0.10}_{-0.24}$ & $0.71^{+0.14}_{-0.16}$ & $0.266^{+0.069}_{-0.253}$ & $0.67^{+0.15}_{-0.16}$ \\ [3pt]
		$10^7\Gamma_0 $ [1/Mpc] & $2.95^{+0.63}_{-2.95}$ & $2.68^{+0.60}_{-2.68}$ & $6.6^{+2.8}_{-4.3}$ & $5.7^{+2.4}_{-3.4}$  \\ [3pt]
		$\log_{10} (\zt)$  & $4.35^{+0.17}_{-0.11}$ & $4.290^{+0.118}_{-0.081}$ & $4.381^{+0.167}_{-0.095}$ & $4.322^{+0.106}_{-0.088}$ \\ [3pt]
		$\sigma 8$ & $0.800^{+0.020}_{-0.013}$ & $0.813^{+0.020}_{-0.013}$ & $0.779^{+0.015}_{-0.017}$ & $0.791^{+0.016}_{-0.016}$  \\ [3pt]
		$\Omega_m$ & $0.3078^{+0.0066}_{-0.0065}$ & $0.2989^{+0.0060}_{-0.0055}$ & $0.3074^{+0.0066}_{-0.0064}$ & $0.2983^{+0.0057}_{-0.0059}$  \\ \hline  &&&&\\[-8pt]		
		$M_b $ & ${-19.368}^{+0.026}_{-0.038}$ & ${-19.291}^{+0.023}_{-0.025}$ & ${-19.369}^{+0.025}_{-0.040}$ & ${-19.294}^{+0.023}_{-0.024}$ \\ [3pt]
		$H_0$ [km/s/Mpc] & $69.29^{+0.93}_{-1.21}$ & $71.92^{+0.78}_{-0.87}$ & $69.27^{+0.84}_{-1.31}$ & $71.86^{+0.79}_{-0.83}$  \\ [3pt]
		$S_8$ & $0.810^{+0.020}_{-0.016}$ & $0.812^{+0.020}_{-0.017}$ & $0.788^{+0.013}_{-0.013}$ & $0.7889^{+0.0014}_{-0.0013}$ \\  [3pt] \hline
		
	\end{tabular}
	\caption{Mean and $\pm 1 \sigma$ values for a fits of WZDR+.}
	\label{tb:mean_and_bf_WZDRplus}
\end{table*}

\begin{table*}[h!]{ }
	\centering
	\begin{tabular}{|c | c | c | c | c |} 
		\hline
		& $\mathcal{D}$ &  $\mathcal{D}$H & $\mathcal{D}$S & $\mathcal{D}$HS  \\
		\hline &&&&\\[-8pt]
		100$\theta_s$ & $1.04251$& $1.04321$ & $1.04258$ & $1.04347$  \\  [3pt]
		$\Omega_b h^2$ &$0.02255$& $0.00227$ & $0.02260$ & $0.02286$ \\ [3pt]
		$ \Omega_{\rm dm}h^2$ & $0.1243$& $0.1303$ & $0.1228$ & $0.1307$\\ [3pt]
		$\ln 10^{10} A_s$ & $3.047$& $3.053$ & $3.052$ & $3.053$ \\ [3pt]
		$n_s$ &$0.9735$& $0.9813$ & $0.9754$ & $0.9867$\\ [3pt]
		$\tau_{\rm reio}$ & $0.0561$& $0.0584$ & $0.0570$ & $0.0574$ \\ [3pt]
		$\nir $ & $0.30$& $0.70$ & $0.21$ & $0.67$ \\ [3pt]
		$10^7\Gamma_0 $ [1/Mpc] & $0.05$& $0.43$ & $4.84$ & $4.99$ \\ [3pt]
		$\log_{10} (\zt)$  & $4.33$& $4.26$ & $4.38$ & $4.32$ \\ [3pt]
		$\sigma 8$ & $0.8203$& $0.8308$ & $0.7836$ & $0.7937$  \\ [3pt]
		$\Omega_m$ & $0.3057$& $0.2980$ & $0.3068$ & $0.2987$ \\ \hline  &&&&\\[-8pt]		
		$M_b $ & $-19.364$& $-19.293$ & $-19.377$ & $-19.294$ \\ [3pt]
		$H_0$ [km/s/Mpc] & $69.46$& $71.83$ & $69.01$ & $71.84$  \\ [3pt]
		$S_8$ & $0.828$& $0.828$ & $0.792$ & $0.792$\\  [3pt] \hline
		\hline \hline &&&&\\[-8pt]
		$\chi^2_{\rm CMB}$ & $2766.15$ & $2769.30$ & $2768.20$ &  $2771.55$  \\ [3pt]
		$\chi^2_{\rm Pantheon}$ & $1025.72$ & $1025.71$ & $1025.75$ & $1025.64$ \\ [3pt]
		$\chi^2_{\rm BAO}$ & $5.14$ & $5.81$ & $5.18$ &  $5.67$ \\ [3pt]
		$\chi^2_{\rm Pl. \ lensing}$  & $(9.30)$ & $(9.75)$  & $9.81$ & $10.43$ \\ [3pt]
		$\chi^2_{\rm S8}$  & $(13.77)$ & $(13.82)$ & $2.19$ & $2.12$  \\ [3pt]
		$\chi^2_{\rm SH0ES}$  & $(16.81)$ & $2.20$ & $(21.22)$ & $2.34$ \\ [3pt]
		\hline &&&&\\[-8pt]
		$\chi^2_{\rm tot}$ & $3797.01$ & $3803.02$ & $3811.13$ & $3817.76$ \\ [3pt]
		\hline
		
	\end{tabular}
	\caption{Best-fit values for a fits of WZDR+.}
	\label{tb:bestfit_WZDRplus}
\end{table*}

\begin{table*}[h!]{ }
	\centering
	\begin{tabular}{|c || c | c | c| c ||c | c | c| c ||c | c | c | c |} 
		\hline  
		& \multicolumn{4}{c||}{$\Lambda{\rm CDM}$} &
		\multicolumn{4}{c||}{SIDR+} &
		\multicolumn{4}{c|}{WZDR+} \\ 
		\hline
		& H0 &  $S_8$ & $\chi^2_{\rm min}$ &$Q_{DMAP}$ & H0 &  $S_8$ & $\chi^2_{\rm min}$&$Q_{DMAP}$& H0 &  $S_8$ & $\chi^2_{\rm min}$&$Q_{DMAP}$\\
		\hline\hline &&& &&&&&&&&&\\[-8pt]
		\dataD & $67.64$& $0.823$ & $3798.66$& &$68.15$& $0.820$ & $3798.33$& &$69.46$& $0.828$ & $3797.01$&\\  [3pt]
		\dataDH & $ 68.63$ & $0.802$ & $3829.68$ &$5.57\,\sigma$&  $ 71.74$ & $0.808$ & $3808.46$&$3.18\,\sigma$& $ 71.83$ & $0.828$ & $3803.03$&2.45$\,\sigma$ \\ [3pt]
		\dataDS& $68.32$ & $0.802$ & $3805.45$ &$2.61\,\sigma$& $69.25$ & $0.794$ & $3806.10$&$2.79\,\sigma$& $69.49$ & $0.787$ & $3801.26$&2.06$\,\sigma$ \\ [3pt]
		\dataDHS& $68.96$ & $0.790$ & $3832.32$ &$5.80\,\sigma$& $72.03$ & $0.791$ & $3811.47$&$3.62\,\sigma$& $71.86$ & $0.790$ & $3807.23$&3.20$\,\sigma$ \\ [3pt]
		\hline
	\end{tabular}
	\caption{Best-fit points used for calculating the $Q_{DMAP}$ values. Note that for the datasets \dataDS and \dataDHS in this table and in the resulting $Q_{DMAP}$ values only we excluded the Planck Lensing likelihood from the $\mathcal{S}$ dataset. This was done because it is not straightforward to define $Q_{DMAP}$ with it included in $\mathcal{S}$. Here $Q_{DMAP}(\mathcal{DX})=\left(\chi^2_\mathcal{DX}-\chi^2_\mathcal{D}\right)^{1/2}$ for $\mathcal{X} = \mathcal{H}, \mathcal{S}, \mathcal{HS}$.}
	\label{tb:bf_for_qdmap}
\end{table*}

\end{document}